\documentstyle[preprint,pra,aps,eqsecnum,amsfonts]{revtex}

\newtheorem{theorem}{Theorem}
\newtheorem{lemma}{Lemma}

\begin{document}

\draft

\title{Boundary Conditions on Internal Three-Body Wave Functions}
\author{Kevin A. Mitchell and Robert G. Littlejohn}
\address{Department of Physics, University of California, 
Berkeley, California 94720}
\date{\today}
\maketitle

\begin{abstract}
For a three-body system, a quantum wave function $\Psi^\ell_m$ with
definite $\ell$ and $m$ quantum numbers may be expressed in terms of
an internal wave function $\chi^\ell_k$ which is a function of three
internal coordinates.  This article provides necessary and sufficient
constraints on $\chi^\ell_k$ to ensure that the external wave function
$\Psi^\ell_m$ is analytic.  These constraints effectively amount to
boundary conditions on $\chi^\ell_k$ and its derivatives at the
boundary of the internal space.  Such conditions find similarities in
the (planar) two-body problem where the wave function (to lowest
order) has the form $r^{|m|}$ at the origin.  We expect the boundary
conditions to prove useful for constructing singularity free
three-body basis sets for the case of nonvanishing angular momentum.
\end{abstract}


\section{Introduction}

Consider a wave function $\Psi^\ell_m$ for a system of three bodies
that is an eigenfunction of $L^2_s$ and $L_{sz}$ with
quantum numbers $\ell$ and $m$, respectively, where ${\bf L}_s$ is the
space-fixed orbital angular momentum.  (The $s$ subscript stands for
``space-fixed''.)  We regard $\Psi^\ell_m$, the ``external wave
function,'' as a function of the two Cartesian Jacobi vectors.  It is
well known \cite{Sutcliffe80,Ezra82,Littlejohn97} that such a wave
function can be written in the form

\begin{equation}
\Psi^\ell_m = \sum_k {\cal D}^{\ell*}_{mk}
\chi^\ell_k,
\label{r31}
\end{equation}
where the Wigner rotation matrix ${\cal D}^{\ell}_{mk}$ is a function of the
three Euler angles and where $\chi^\ell_k$, the ``internal wave
function,'' is a function of three internal or shape coordinates. As
usual, $k$ is regarded as the quantum number of the body-fixed $L_z$.
The wave function $\Psi^\ell_m$ need not be an energy eigenfunction;
for example, it could be an element of a basis set in terms of which
an unknown energy eigenfunction is to be expanded.  The basis sets we
have in mind include standard orthonormal bases, hyperspherical
harmonics \cite{Aquilanti86a,Aquilanti98}, discrete variable
representation (DVR) bases
\cite{Harris65,Dickinson68,Lill82,Light85,Lill86,Baye86}, and wave
packet \cite{Davis79} or wavelet \cite{Daubechies92} bases.

This paper concerns boundary conditions which the internal
wave function $\chi^\ell_k$ must satisfy, given that the external
wave function $\Psi^\ell_m$ is a smooth function of the
Cartesian Jacobi vectors.  The boundary  in question is
 the boundary of shape space, which consists of the collinear
configurations.  The applications we have in mind are
mainly molecular (either bound states of triatomic molecules or
triatomic scattering problems), but the considerations we raise also
apply to atoms or other systems of bodies with Coulomb interactions
(with certain qualifications discussed below).  We ignore spin in this
paper.  We consider only three-body systems in this paper, but an
important reason for studying boundary conditions in three-body
systems is that it is good practice for the analogous problem for
systems of four or more bodies, which is generally more difficult
and much less well understood.

There are at least two practical reasons for being interested in
boundary conditions.  First, if one attempts to expand some unknown
function in terms of a given basis, and if there are boundary
conditions satisfied by the unknown function which are not satisfied
by the basis functions, then the convergence will be slow.  In
important cases, the coefficient of the $n$-th term in the expansion
will fall off either exponentially or algebraically with $n$,
depending on whether the basis does or does not satisfy the required
boundary conditions, respectively.  For example, it is a bad idea to
use the ordinary Legendre polynomials to expand a function whose
$\theta$ dependence has the boundary conditions of one of the
associated Legendre functions for $m>0$.  The importance of properly
treating such boundary conditions in three-body problems has been
discussed previously by Kendrick et. al. \cite{Kendrick99}.

Sometimes basis functions are created on the internal space simply by
writing down some internal wave functions $\chi^\ell_k$ that are
considered convenient, for example, distributed Gaussians or wave
packets.  From the given internal wave functions, one can construct
the corresponding external wave functions according to
Eq.~(\ref{r31}).  The question then arises, will these external wave
functions have the same smoothness and analyticity properties as some
unknown wave function (usually an energy eigenfunction) which one
wishes to find?  If not, the convergence will be poor.  For another
example, it is common practice to create internal basis functions by
writing down the exact internal Hamiltonian, and then carving out some
piece of it which has eigenfunctions which can be determined
analytically.  Again there is a question as to whether the basis
functions created in this manner have the boundary conditions required
of the desired unknown eigenfunction.  The answer to this question
depends in part on whether the operator created by carving out a piece
of the Hamiltonian is itself well behaved.  The analysis of this paper
will show how to answer these questions.

A second reason for being interested in boundary conditions is that
numerical methods for solving partial differential equations on a grid
must generally take careful account of boundary conditions, in order
to guarantee reasonable accuracy and convergence.  Grid and basis set
methods are related, since grid methods implicitly involve a set of
localized basis functions associated with the grid points.  For
example, DVR methods involve a basis set consisting of localized wave
functions, resembling diffraction patterns from a narrow slit.  The
often cited ``unexpected accuracy'' (that is, rapid convergence) of
DVR methods is closely related to satisfying the right boundary
conditions.  For example, the trapezoidal rule converges exponentially
(rapidly) when an analytic, periodic function is integrated over a
period, but has only power law (slow) convergence if the function is
analytic but not periodic, or if it is integrated only over a partial
period.  Similarly, if the wrong DVR basis set is used for a given
problem, the (by now expected) unexpected accuracy will be lost.  We
will have more to say about boundary conditions and rates of
convergence in future publications, but this paper will concentrate on
the boundary conditions themselves.

We will momentarily present the principal results of our analysis of
boundary conditions in three-body systems, but first it is well to
recall some facts about two-body systems with rotational
invariance.  Thinking of energy eigenfunctions, we will speak first of
the problem of central force motion.  In three spatial dimensions, the
energy eigenfunction can be written $\Psi^\ell_m({\bf r}) =
Y^\ell_m(\theta,\phi) \chi_\ell(r)$, where $\chi_\ell(r)$ is the
internal or radial wave function, defined on the radial half-line
$0\le r<\infty$, which is the internal space.  According to the
standard textbook analysis \cite{Messiah66}, the radial wave function behaves
as $r^\ell$ near $r=0$.  This behavior holds when the true potential
$V(r)$ is analytic at $r=0$, but also in other cases such as that of
the singular Coulomb potential.

The standard analysis that produces these results proceeds by
expanding the radial wave function in a Taylor series about $r=0$ and
balancing terms on the two sides of the Schr\"odinger equation.
Unfortunately, this analysis leaves the impression that the behavior
$\chi_\ell \sim r^\ell$ of the radial wave function near $r=0$ applies
only to energy eigenfunctions.  Actually this behavior is much more
general.  Consider any wave function $\Psi^\ell_m({\bf r})$ which is
an eigenfunction of $L^2$ and $L_z$ and which is analytic at ${\bf
r}=0$ when expressed in terms of the Cartesian coordinates $(x,y,z)$.
This would apply to the eigenfunctions of any rotationally invariant
operator which is well behaved at $r=0$, including Hamiltonians with
central potentials $V(r)$ which are analytic at $r=0$.  With standard
assumptions about phase conventions the wave function can be written
$\Psi^\ell_m({\bf r}) = Y^\ell_m(\theta,\phi) \chi_\ell(r)$, where
$\chi_\ell(r)$ is the radial wave function. We note that this form
follows from the standard theory of rotations and the fact that
$\Psi^\ell_m$ is an eigenfunction of $L^2_s$ and $L_{sz}$; we do not invoke
separation of variables, since we are not necessarily separating any
wave equation.  Then it turns out that $\chi_\ell(r)$ is analytic at
$r=0$, and that its Taylor series begins with the $r^\ell$ term and
thereafter contains only the powers $r^{\ell+2n}$, $n=1,2,\ldots$,
that is, every other integer power of $r$.  Energy eigenfunctions in
the Coulomb problem do not fit this pattern, since $\Psi^\ell_m({\bf
r})$ has a cusp at $r=0$ and is not analytic there.  This is because
$V(r)$ is not analytic at $r=0$.  Although Coulomb radial wave
functions $\chi_\ell$ do go as $r^\ell$ near $r=0$, the Taylor series
of $\chi_\ell$ contains every subsequent power of $r$, not every other
one.

It is also worthwhile mentioning the case of two bodies in two
spatial dimensions, since planar two-body boundary conditions bear a
strong analogy to the boundary conditions in three-body systems.
For the planar problem, an eigenfunction of $L_{sz}$ can be written
$\Psi_m({\bf r})=e^{im\phi} \chi_m(r)$, where ${\bf r}=(x,y)$, $\phi$
is the azimuthal angle, and $m=0, \pm1, \pm2,
\ldots$ is the quantum number of $L_{sz}$.  If $\Psi_m$ is analytic
in the two Cartesian coordinates $(x,y)$ at ${\bf r}=0$, then the
radial wave function $\chi_m(r)$ is analytic at $r=0$, its Taylor
series begins with the $r^{|m|}$ term, and subsequently contains only
every other power of $r$, $r^{|m|+2n}$, $n=1,2,\ldots$.

At this point a reader who works with three-body molecular problems
may wonder what the relevance is of two-body boundary conditions at
$r=0$, since $r=0$ is the two-body collision and the collision of
atoms in molecular problems does not happen at ordinary energies.  The
answer is that in three-body systems the collinear configurations play
somewhat the same role as the two-body collision in the planar
two-body problem, insofar as boundary conditions are concerned.  This
is why it is important to know about two-body boundary conditions at
$r=0$, even for molecular problems.  Collinear configurations are not
necessarily suppressed in three-body molecular problems, and are often
important.

On the other hand, collisional configurations are important in Coulomb
problems, where the wave function has cusp-like singularities
\cite{Kato57,Fock54}, again because of the nonanalyticity of the
potential.  Since this paper studies the boundary conditions on the
internal wave function which result from the analyticity of the
external wave function, and since the external wave function in
Coulomb problems is not analytic at collisions, the analysis of this
paper does not apply to collisional configurations in Coulomb
problems. But of course Coulomb problems also have collinear
configurations, and our analysis does apply to these, as long as they
are not also collisional.  We can summarize by saying that the
analysis of this paper applies to all the important boundary
conditions in three-body molecular problems, and to some of them (the
collinear, noncollisional configurations) in Coulomb problems.

We now return to the three-body problem, and summarize the main
results of this paper, which relate the analyticity of the external
wave function $\Psi^\ell_m$ to the behavior of the internal wave
function $\chi^\ell_k$ at the boundary of the internal space.  The
boundary of collinear shapes is a plane specified by $\Theta=0$ in
Smith's \cite{Smith62} hyperspherical coordinates, or $w_3=0$ in the
coordinates to be introduced below.  One point of this plane is the
three-body collision, which is excluded from our analysis.  At other
points of this plane, we have established necessary and sufficient
conditions on $\chi^\ell_k$ such that the external wave function
$\Psi^\ell_m$ should be analytic functions of the six Cartesian
components of the two Jacobi vectors.  If $\Psi^\ell_m$ is analytic at
a collinear configuration (not the three-body collision), then the
internal wave function $\chi^\ell_k$ satisfies the following
properties.  First, $\chi^\ell_k$ is itself analytic at the collinear
shape, when expressed in terms of certain internal coordinates to be
described below.  Suffice it for now to say that one of these
coordinates, call it $w_3$, measures the mass-weighted distance from
the plane $w_3=0$ of collinear shapes, while the other two
coordinates, $w_1$ and $w_2$, indicate where we are on this plane.

The second property involves a modified version of the internal wave
function $\chi^\ell_\mu$ (in contrast to $\chi^\ell_k$).  The
distinction is that $k$ is the eigenvalue of body-fixed $L_z$, whereas
$\mu$ is the eigenvalue of body-fixed $\hat {\bf n} \cdot {\bf L}$,
where $\hat {\bf n}$ is the body-referred unit vector specifying the
axis of a collinear shape.  The vector $\hat{\bf n}$ is defined on the
boundary plane of the internal space (excluding the triple collision),
and is a function of where we are on that plane.  Both $k$ and $\mu$
range from $-\ell$ to $+\ell$, and the two internal wave functions are
related by a rotation which maps the body $\hat {\bf z}$-axis into the
$\hat {\bf n}$-axis.  Then, as we shall show, it turns out that if
$\chi^\ell_\mu$ is expanded in powers of $w_3$, corresponding to
movement in the internal space away from the boundary in the direction
of increasing $w_3$, then the first nonvanishing power is
$w_3^{|\mu|}$, and subsequently only every other power occurs in the
Taylor series, $w_3^{|\mu|+2n}$, $n=1,2,\ldots$.

The converse is also true: if $\chi^\ell_k$ satisfies these two
properties at a collinear shape (not the triple collision), then the
external wave function $\Psi^\ell_m$ is analytic.  Obviously, these
boundary conditions are like those of the planar two-body  problem at
$r=0$, with $w_3$ playing the role of $r$.  

We have also proved the (plausible) fact that at configurations which
are not at the boundary of the internal space (noncollinear
configurations), the external wave function $\Psi^\ell_m$ is analytic if
and only if $\chi^\ell_k$ is analytic.  The stated conditions, both on
and off the boundary plane, are independent of the convention for
Euler angles or the convention for body frame, assuming we avoid
certain body frame singularities \cite{Pack94,Littlejohn98a,Littlejohn98b}. 

We exclude the triple collision because there is an inevitable
``string'' singularity \cite{Littlejohn98a} of the body frame in the
neighborhood of this configuration that complicates the analysis.
This configuration is not important in three-body molecular problems,
but is so for Coulomb problems (where the singularity of the potential
is a further complication).  

The outline of this paper is as follows.  Section~\ref{s2} contains
the main result of the paper, Theorem~\ref{t1}, which concerns the
boundary conditions satisfied by the internal wave function.  We have
stated this theorem in as nontechnical language as possible.
Section~\ref{s2} also contains a description of how the boundary
conditions are to be applied and discusses several explicit
conventions for internal coordinates and body frame.
Sections~\ref{s3} and \ref{s4} are more technical, and are devoted to
proving Theorem~\ref{t1}.  Section~\ref{s3} states and proves boundary
conditions for the simpler case of the planar two-body problem.
Section~\ref{s4} uses the results of Sect.~\ref{s3} to prove the
conditions for the three-body problem.  Section~\ref{s5} contains the
conclusions.  An Appendix collecting several facts about the
representations of $SO(2)$ is included for reference.

\section{The Three-Body Boundary Conditions}
\label{s2}

\subsection{Review of three-body formalism}

Before stating the boundary conditions on three-body wave functions,
we review some necessary facts about the three-body problem.  We
closely follow the notation and spirit of Littlejohn and Reinsch
\cite{Littlejohn97}. A three-body configuration in the center of mass
frame is described by two (mass-weighted) Jacobi vectors ${\bf r}_{s
\alpha}$, $\alpha = 1,2$.  Jacobi vectors are a standard topic in $n$-body
problems \cite{Aquilanti86,Littlejohn97}.  The $s$ subscript indicates
that ${\bf r}_{s \alpha}$ is referred to the space, or laboratory,
frame.  For obvious reasons, we will call the space of Jacobi vectors
``configuration space''.

It is often convenient to specify a configuration by its shape and
orientation.  By the shape, we mean the configuration modulo physical
rotations; it is parameterized by three rotationally invariant
quantities (called internal, or shape, coordinates).  We will denote
shape coordinates in general by $q_\mu, \mu = 1,2,3$.  A specific and
particularly useful set of such coordinates is $(w_1, w_2, w_3)$
(henceforth called the ``$w$-coordinates'') defined by

\begin{eqnarray}
w_1 & = & r_{s1}^2 - r_{s2}^2
=  \rho^2 \cos 2\Theta \cos 2 \Phi, 
\label{r32} \\
w_2 & = & 2 {\bf r}_{s1} \cdot {\bf r}_{s2}
= \rho^2 \cos 2 \Theta \sin 2 \Phi, 
\label{r33} \\
w_3 & = & 2 |{\bf r}_{s1} \bbox{\times} {\bf r}_{s2}|
= \rho^2 \sin 2 \Theta,
\label{r34}
\end{eqnarray}
where we have expressed the $w$-coordinates both in terms of the
Jacobi vectors and in terms of Smith's \cite{Smith62,Aquilanti93} symmetric
hyperspherical coordinates $(\rho,\Theta,\Phi)$.  Here $\rho =
(r_{s1}^2 + r_{s2}^2)^{1/2}$ is the hyperradius.  The $w$-coordinates
have been used by many researchers over the years, including Gronwall
\cite{Gronwall32}, Smith\cite{Smith62}, Dragt\cite{Dragt65},
Iwai\cite{Iwai87b}, Aquilanti et. al. \cite{Aquilanti93}, and others.
The $w$-coordinates have ranges $-\infty < w_1, w_2 < \infty$, $0 \le
w_3 <\infty$ and are in one-to-one correspondence with three-body
shapes.  Thus shape space is the closed half-space containing the
physical region of coordinate space $w_3 \ge 0$.  Sometimes it will
also be convenient to consider the region of coordinate space $w_3 <
0$, which we call the unphysical region. The boundary plane $w_3 = 0$
consists of all collinear shapes.  The boundary conditions to be
presented below occur along this plane.  We define $w = (w_1^2 + w_2^2
+ w_3^2)^{1/2}$ to be the ``radius'' in $w_1 w_2 w_3$-space and
note the identity,

\begin{equation}
 w = (w_1^2 +w_2^2 +w_3^2)^{1/2} = \rho^2.
\end{equation}
The origin $w = 0$ is the three-body collision and is an
especially singular point.

The orientation of a configuration is defined relative to some
convention for body frame.  A body frame convention may be defined by
specifying the functions ${\bf r}_\alpha(q)$, where
${\bf r}_\alpha$ represents the body components of the Jacobi vectors
and where $q$ represents an arbitrary shape.  (Equivalently, $q$
stands for $(q_1, q_2, q_3)$).  Our rule is to attach an $s$-subscript
to quantities referred to the space frame, and to omit this subscript
for quantities referred to the body frame.  The quantities ${\bf
r}_\alpha(q)$ can also be thought of as specifying a reference
orientation for a given shape $q$, relative to which other
orientations of the same shape are referred.  In the reference
orientation, the space frame is identical to the body frame.  The
orientation of the configuration is defined as the rotation matrix
${\sf R} \in SO(3)$ which rotates the reference configuration into the
actual configuration,

\begin{equation}
{\bf r}_{s \alpha} = {\sf R} {\bf r}_\alpha(q).
\label{r20}
\end{equation}
We discuss several common choices of body frame in Sect.~\ref{s6}, but
for now we leave the body frame unspecified.

For collinear shapes, ${\sf R}$ is not uniquely determined by
Eq.~(\ref{r20}) and there is no unique body frame associated with a
particular choice of reference orientation.  Nevertheless, the
functions ${\bf r}_\alpha(q)$ are normally well defined at collinear
shapes, and the assignment of a reference orientation for a collinear
shape will prove to be a useful concept.  We will have more to say in
Sect.~\ref{s7} about the singular nature of the body frame at
collinear shapes.  

We now turn to the three-body wave function $\Psi$ and give a quick
derivation of Eq.~(\ref{r31}) from angular momentum theory.  The wave
function $\Psi$ depends on the two Jacobi vectors ${\bf r}_{s\alpha}$.
Rotations act on such wave functions by

\begin{equation}
\bbox{(}{\cal R}({\sf Q}) \Psi \bbox{)} ({\bf r}_{s \alpha}) 
= \Psi({\sf Q}^T {\bf r}_{s \alpha}),
\label{r36}
\end{equation}
where ${\cal R}$ is the rotation operator parameterized by the
rotation matrix ${\sf Q} \in SO(3)$, and $T$ is the transpose.  We
consider a collection of $2\ell + 1$ wave functions $\Psi^\ell_{m}$,
$-\ell \le m \le \ell$, with definite total angular momentum $\ell$
and transforming under the action of $SO(3)$ via the standard
representation,

\begin{equation}
\Psi^\ell_m({\sf Q}^T{\bf r}_{s \alpha}) 
= \bbox{(} {\cal R}({\sf Q}) \Psi^\ell_{m} \bbox{)} ({\bf r}_{s \alpha}) 
= \sum_k  {\cal D}^\ell_{k m}({\sf Q}) \Psi^\ell_{k}({\bf r}_{s \alpha}),
\label{r5}
\end{equation}
where ${\cal D}^\ell_{k m}({\sf Q})$ is the Wigner rotation matrix of
${\sf Q}$.  We use the (active) conventions of Messiah
\cite{Messiah66} and Biedenharn and Louck\cite{Biedenharn81a}.  We
define the internal wave function by

\begin{equation}
\chi^\ell_k (q) = \Psi^\ell_k \bbox{(} {\bf r}_\alpha(q) \bbox{)}.
\label{r6}
\end{equation}
The internal wave function is a multicomponent wave function which we
call a ``spinor.''  The spinor index is $k$, $-\ell \le k \le \ell$.
Using Eqs.~(\ref{r20}), (\ref{r5}), and (\ref{r6}) we obtain the final result

\begin{equation}
\Psi^\ell_m({\bf r}_{s\alpha}) 
= \Psi^\ell_m \bbox{(} {\sf R}{\bf r}_\alpha(q) \bbox{)}
= \sum_k {\cal D}^{\ell *}_{ m k}({\sf R})\chi^\ell_k(q).
\label{r7}
\end{equation}
The importance of Eq.~(\ref{r7}) is that it completes the one-to-one
correspondence between the external and internal wave functions.
Equation~(\ref{r6}) gives the internal wave function in terms of the
external wave function while Eq.~(\ref{r7}) gives the external wave
function in terms of the internal wave function.  One must realize,
however, that all $2 \ell + 1$ components of $\chi^\ell_k$ must be
specified to construct $\Psi^\ell_m$, whereas only one $\Psi^\ell_m$
is needed to construct $\chi^\ell_k$.  This is because the different
$\Psi^\ell_m$'s are not independent but related by raising and
lowering operations, whereas the $\chi^\ell_k$'s are not.

\subsection{Statement of Boundary Conditions}

\label{s7}

We now turn to the principal question addressed in this paper: What
are necessary and sufficient conditions on the internal wave function
$\chi^\ell_k(q)$ which ensure that the external wave function
$\Psi^\ell_m({\bf r}_{s \alpha})$ is analytic?  Naively, one might
expect the only condition to be that $\chi^\ell_k(q)$ is itself
analytic, but this is not sufficient as we now explain.  

First, we review some fundamental issues regarding analyticity.
Further clarification of these points may be found in any basic
reference on differential geometry, for example
Refs.~\cite{Spivak79,Nakahara90}.  To say that a function of several
variables is analytic at a point means that the function agrees with
its Taylor series in a neighborhood of that point. To say that a
function defined on a smooth manifold is analytic at a point means
that the function, when represented in a suitable choice of
coordinates, is an analytic function of those coordinates at that
point.  The choice of coordinates is critical since a function
analytic with respect to one set of coordinates may not be analytic
with respect to another.  Thus, whenever we say that a function
defined on a smooth manifold is analytic, we must be careful to say
with respect to what set of coordinates.  Now, if two sets of
coordinates are related to one another by an invertible analytic
function (with analytic inverse) then these two sets of coordinates
are said (in standard mathematical terminology) to be ``compatible.''
A function analytic with respect to one set of coordinates is also
analytic with respect to a compatible set of coordinates.  Thus, the
analyticity of a function is defined relative to an entire set of
compatible coordinates.

As an example, consider the plane ${\Bbb R}^2$, and take the standard
$(x,y)$ coordinates as the privileged coordinates defining
analyticity.  Polar coordinates $(r,\theta)$ are compatible with
$(x,y)$ everywhere except at the origin (and on a radial line).  Thus,
a function such as $f(r,\theta)=r$ which is an analytic function of
the polar coordinates may still not be an analytic function at the
origin of ${\Bbb R}^2$.  One needs additional boundary conditions at
the origin to guarantee that a function analytic in polar coordinates
is truly analytic on ${\Bbb R}^2$.  This example contains the core
idea of why boundary conditions may be needed to guarantee analyticity
of a function.  We explore this example further in Sect.~\ref{s3}.

When considering the three-body wave function $\Psi^\ell_m$, we take
the privileged set of coordinates defining analyticity to be the
Jacobi coordinates $r_{s \alpha k}$, $\alpha=1,2$, $k=x,y,z$.  The
reason for choosing these coordinates is that in practice the
potential energy and the wave functions are typically analytic (except
at collisions) with respect to these coordinates.  So long as one only
uses Jacobi coordinates, this fact is sufficient and the rest of this
paper could be skipped.  However, as already pointed out, a shape and
orientation description of configuration space is often advantageous
and this naturally involves expressing $\Psi^\ell_m$ in shape and
orientation coordinates as in Eq.~(\ref{r7}).  An important fact is
that shape and orientation coordinates are never compatible at
collinear shapes, because the rotation matrix ${\sf R}$ is not defined
by Eq.~(\ref{r20}).  This is much like the relation between
rectangular and polar coordinates at the origin of the plane,
discussed earlier, and it explains why the analyticity of
$\chi^\ell_k$ alone is not sufficient to guarantee the analyticity of
$\Psi^\ell_m$ at collinear shapes.

We will also be interested in the analyticity of functions
defined on the internal space (such as the internal wave function
$\chi^\ell_k$).  We have found it most useful to take the
$w$-coordinates as the privileged coordinate system with respect to
which the analyticity of such functions is defined.  All internal
coordinate systems in common use are compatible with the
$w$-coordinates at most locations in shape space.  On the internal
space there is the additional issue of what we mean by analyticity at
the boundary $w_3=0$.  We will say that a function defined on the
internal space is analytic at a particular point on the boundary if
the function has an analytic continuation into the nonphysical region
$w_3<0$ in the neighborhood of that point.

The analyticity of the functions ${\bf r}_\alpha(q)$ requires special
comment.  These functions have singularities on certain curves
(``strings'') in $w_1 w_2 w_3$-space, which emanate from the three-body
collision $w=0$ and go out to infinity.  The location of these strings
depends on the convention for body frame
\cite{Pack94,Littlejohn98a,Littlejohn98b}.  In the following we wish
to work with body frame conventions which remove the strings from the
region of interest in the internal space.

The reader may wonder about the analyticity of the Euler angles, or of
functions of them.  As it turns out, we never need to worry about such
issues, because the main result of this paper, which is the
establishment of necessary and sufficient conditions for the
analyticity of $\Psi^\ell_m$, only involves the analytic properties of
$\chi^\ell_k$.  The results we prove below are valid for arbitrary
conventions for Euler angles.

Turning away now from general notions of analyticity, we observe that
at a collinear shape, which is not the triple collision, there is a
well-defined (up to sign) unit vector, denoted by $\hat{\bf n}$,
pointing along the body-referred axis of collinearity.  The vector
$\hat{\bf n}$ depends on the position along the boundary plane and is
undefined off of the plane.  We therefore choose a convention for extending
$\hat{\bf n}$ off of the boundary plane.  That is, we choose a
function $\hat{\bf n}(q)$, defined on shape space, which points along
the collinear axis when evaluated at a collinear shape.  When
evaluated at a noncollinear shape, we only require that $\hat{\bf
n}(q)$ lie in the plane spanned by ${\bf r}_1(q)$ and ${\bf r}_2(q)$.

We now introduce a certain
basis of spinors $\tau_\mu(q)$, $-\ell \le \mu \le \ell$, which are
eigenspinors of the projection $\hat{\bf n}(q) \cdot {\bf L}$ of the
body-referred angular momentum operator ${\bf L}$ onto the collinear axis.
These spinors are chosen to satisfy

\begin{eqnarray}
(\hat{\bf n} \cdot {\bf L}) \tau_\mu & = & \mu \tau_\mu,
\label{r42} \\
\tau_\mu^\dagger \tau_{\mu'} 
& = & \sum_k (\tau_\mu)^*_k (\tau_{\mu'})_k 
=  \delta_{\mu \mu'},
\label{r78}
\end{eqnarray} 
where $\dagger$ represents the Hermitian conjugate.  Here the $\mu$
subscript does not index the components of $\tau_\mu$ but rather
labels the spinor; the $2\ell + 1$ components themselves are denoted
by $(\tau_\mu)_k$, $-\ell \le k \le \ell$ and are taken with respect
to the normalized eigenbasis of $L_z$, using standard phase
conventions.  Notice that Eqs.~(\ref{r42}) and (\ref{r78}) determine
$\tau_\mu(q)$ only up to an overall phase, which is another convention
we have the freedom to choose.  We define an alternative version of
the internal wave function by

\begin{equation}
\chi^\ell_\mu = \tau_\mu^\dagger \chi^\ell = \sum_k (\tau_\mu)^*_k \chi^\ell_k,
\label{r46}
\end{equation}
where $\chi^\ell$ is the column vector $(\chi^\ell_{-\ell}, ...,
\chi^\ell_\ell)^T$.  Since the spinors $\tau_\mu$ are orthonormal,
Eq.~(\ref{r46}) may be inverted to give

\begin{equation}
\chi^\ell_k =  \sum_\mu (\tau_\mu)_k \chi^\ell_\mu.
\label{r79}
\end{equation}

Given some region of interest in the internal space, there are
conventions involved in choosing the coordinates $q_\mu$, the
functions ${\bf r}_\alpha(q)$, which specify the reference
orientations, the function $\hat{\bf n}(q)$, which extends the
collinear axis away from the boundary, and the spinors $\tau_\mu(q)$.
We will be particularly interested in a certain class of conventions
which taken together we call ``valid'' conventions.  At a noncollinear
shape, a set of conventions is said to be valid if $q_\mu$ is
compatible with the $w$-coordinates and the functions ${\bf
r}_\alpha(q)$, $\hat{\bf n}(q)$, and $(\tau_\mu)_k(q)$ are analytic.
In particular, this means that the shape in question does not lie on a
string singularity.  At a collinear shape, we still require the
compatibility of the internal coordinates with the $w$-coordinates and
the analyticity of ${\bf r}_\alpha(q)$, $\hat{\bf n}(q)$, and
$(\tau_\mu)_k(q)$.  However, we further require that the boundary
plane be given by $q_3=0$.  We also require that various functions be
either even or odd. Specifically,

\begin{eqnarray}
q_\mu(-w_3) & = & q_\mu(w_3), \hskip 1cm \mu = 1,2,
\label{r70} \\
q_3(-w_3) & = & -q_3(w_3), 
\label{r71} \\
(\hat{\bf n} \cdot {\bf r}_\alpha)(-q_3) 
& = & (\hat{\bf n} \cdot {\bf r}_\alpha)(q_3), 
\label{r72} \\
({\sf P}_\perp {\bf r}_\alpha)(-q_3) 
& = & - ({\sf P}_\perp {\bf r}_\alpha)(q_3),
\label{r73} \\
\hat{\bf n}(-q_3) & = & \hat{\bf n}(q_3), 
\label{r74} \\
(\tau_\mu)_k(-q_3) & = & (\tau_\mu)_k(q_3),
\label{r101}
\end{eqnarray}
where we have suppressed the dependence on $w_1$, $w_2$, $q_1$, and
$q_2$ and where ${\sf P}_\perp(q)$ denotes the projection operator
onto the plane orthogonal to $\hat{\bf n}(q)$.

With the preceding setup, we state the main result of this paper.

\begin{theorem}
\label{t1}
Let a configuration ${\bf r}_{s \alpha}$ (which is not the triple
collision) have shape $q$, and assume valid conventions for the shape
coordinates $q_\mu$, the reference orientation ${\bf r}_{\alpha}(q)$,
the vector $\hat{\bf n}(q)$, and the spinor $\tau_\mu(q)$ in the
neighborhood of $q$.
\vskip 12pt

\noindent 
(i) If $q$ is noncollinear, then $\Psi^\ell_m$ is analytic at ${\bf
r}_{s \alpha}$ if and only if $\chi^\ell_\mu$ (equivalently
$\chi^\ell_k$) is analytic at $q$.
\vskip 12pt

\noindent (ii) If $q$ is collinear, then $\Psi^\ell_m$ is analytic at
${\bf r}_{s \alpha}$ if and only if $\chi^\ell_\mu$ (equivalently
$\chi^\ell_k$) is analytic at $q$ with the Taylor series

\begin{equation}
\chi^\ell_\mu(q_1, q_2, q_3) 
= \sum_{n = 0}^\infty a_{\mu n}(q_1,q_2) q_3^{|\mu| + 2n},
\label{r12}
\end{equation}
where $a_{\mu n}$ is an analytic function of $(q_1, q_2)$.

\end{theorem}

The most striking aspect of this theorem is that the wave function
grows as $q_3^{|\mu|}$ away from the collinear shapes.  We can provide
the following heuristic physical interpretation of this rule.  If a
classical three-body system, under the influence of a smooth
potential, approaches a collinear configuration, any angular momentum
about the collinear axis will create a centrifugal barrier which
prevents the system from reaching the collinear configuration.
Quantum mechanically, the centrifugal barrier acts to suppress the
wave function in the classically forbidden region.  The more quanta of
angular momentum about the collinear axis, the more the wave function
is suppressed resulting in the $q_3^{|\mu|}$ growth.  This
interpretation is dynamical in nature since it depends on the notion
of a Hamiltonian.  We stress, however, that the derivation of Theorem
\ref{t1} will depend only on notions of analyticity and symmetry.

\subsection{Explicit Examples of Boundary Conditions}

\label{s6}
For concreteness, we analyze the boundary conditions explicitly for
several choices of valid conventions.  The first example uses the
$w$-coordinates and a body frame which coincides with the principal
axes.  The body-referred Jacobi vectors are given parametrically by

\begin{eqnarray}
{\bf r}_1(w_1, w_2, w_3) & = & 
\left[ {a + b\over 2\sqrt{2}} 
\left( {1 + {w_1 \over ab}}\right)^{1/2} {w_2 \over |w_2|} \right]
\hat{\bf z} 
- \left[ { a - b  \over 2 \sqrt{2}}
\left( 1 - {w_1 \over ab} \right)^{1/2}  \right]
\hat{\bf x}, 
\label{r8}\\
{\bf r}_2(w_1, w_2, w_3) & = & 
\left[ { a + b \over 2\sqrt{2}} 
\left(1 - {w_1 \over ab} \right)^{1/2} \right]
\hat{\bf z} 
+  \left[ { a - b \over 2\sqrt{2}}
\left( 1 + {w_1 \over ab}\right)^{1/2} {w_2 \over |w_2|} \right]
\hat{\bf x}, 
\label{r9} \\
a & = & \sqrt{w + w_3}, \\
b & = & \sqrt{w - w_3}.
\end{eqnarray}
Since the collinear axis for a collinear shape is given by $\hat{\bf
n}(w_1, w_2) = \hat{\bf z}$,  we define the extension to
noncollinear shapes by

\begin{equation}
\hat{\bf n}(w_1, w_2, w_3) = \hat{\bf z}.
\label{r75}
\end{equation}
Equations (\ref{r8}) and (\ref{r9}) exhibit a discontinuity in the
function ${\bf r}_\alpha(w_1, w_2, w_3)$ along the $w_3$-axis.  In
fact the reference orientation approaches a continuum of different
values, depending on the direction of approach.  The $w_3$-axis forms
the string singularity of the principal axis frame.  This string
consists of oblate symmetric tops, and the discontinuity there is a
direct result of the ambiguity in the choice of the principal axes due
to the degeneracy of the principal moments of inertia.  Another
singularity occurs in the principal axis frame, arising from the
double-valued nature of the frame.  Upon circling the $w_3$ axis once,
the principal axis frame does not return to its original value, but
rather has rotated by $\pi$.  Thus, to make the principal-axis frame
single-valued requires introducing a branch cut.  In Eqs.~(\ref{r8})
and (\ref{r9}), we have chosen this branch cut to be the
two-dimensional surface $w_2 = 0$, $w_1 > 0$.  Such string
singularities and branch cuts are common to other choices of body
frame as well and are discussed further in Ref.~\cite{Littlejohn98a}.

We comment briefly on the validity of the conventions introduced here.
The $w$-coordinates are trivially valid everywhere, with the $w_3$
coordinate being transverse to the collinear shapes.  The function
$\hat{\bf n}(q) = \hat{\bf z}$ is constant and hence analytic
everywhere.  It obviously satisfies Eq.~(\ref{r74}) as well.  Away
from the frame singularities discussed above, ${\bf r}_\alpha(q)$ is
analytic, and it is straightforward to show that Eqs.~(\ref{r72}) and
(\ref{r73}) are satisfied.  From Eq.~(\ref{r75}) we may take
$(\tau_\mu)_k(q) = \delta_{\mu k}$, which is clearly analytic and
satisfies Eq.~(\ref{r101}).  Hence these conventions are valid
everywhere except at the frame singularities.

Away from the frame singularities and away from the boundary of shape
space, Theorem \ref{t1} tells us that analyticity of $\chi^\ell_k(q)$
is a necessary and sufficient condition for analyticity of
$\Psi^\ell_m({\bf r}_{s \alpha})$.  In order to guarantee analyticity
of $\Psi^\ell_m({\bf r}_{s \alpha})$ on the boundary of shape space
(away from the positive $w_1$-axis where there is a frame
singularity), $\chi^\ell_k(q)$ must have the Taylor series

\begin{equation}
\chi^\ell_k(w_1, w_2, w_3) 
= \sum_{n = 0}^\infty a_{kn}(w_1,w_2) w_3^{|k| + 2n}, 
\label{r11}
\end{equation}
where of course $a_{kn}(w_1, w_2)$ is analytic.

In the next example, we keep the principal axis frame, but change
coordinates from the $w$-coordinates to the symmetric hyperspherical
coordinates $(\rho, \Phi, \Theta)$ defined in Eqs.~(\ref{r32}) --
(\ref{r34}).  The coordinate $\Theta$ is transverse to the boundary of
shape space which occurs at $\Theta = 0$.  These hyperspherical
coordinates are compatible everywhere except the $w_3$-axis and the
two-dimensional surface $w_2 = 0$, $w_1>0$, where they experience a
coordinate singularity.  Note that the location of the coordinate
singularities agrees exactly with the location of the singularities in
the principal axis frame.  We again choose $\hat{\bf n}(q) = \hat{\bf
z}$ and $(\tau_\mu)_k = \delta_{\mu k}$.  It is again true that the
conventions are valid everywhere except at the frame singularities.
Little modification to the form of Eq.~(\ref{r11}) is necessary except
to change the coordinates which gives

\begin{equation}
\chi^\ell_k(\rho, \Phi, \Theta) 
= \sum_{n = 0}^\infty a_{kn}(\rho,\Phi) \Theta^{|k| + 2n}. 
\end{equation}
Of course, the coefficients $a_{kn}(\rho, \Phi)$ in the above equation are
different from those in Eq.~(\ref{r11}).

In the next example, we again use the $w$-coordinates but choose a
different body frame, called the $zxz$-frame in
Ref.~\cite{Littlejohn97} or the $\mbox{BF}_{\tau 1}$ frame in
Ref.~\cite{Pack94}.  This frame places ${\bf r}_1$ along the positive
$z$-axis and ${\bf r}_2$ in the $xz$-plane.  Explicitly, the reference
configuration ${\bf r}_\alpha(w_1, w_2, w_3)$ for the $zxz$-frame is

\begin{eqnarray}
{\bf r}_1(w_1, w_2, w_3) 
& = & {1 \over \sqrt{2}} (w + w_1)^{1/2} \hat{\bf z}, 
\label{r76} \\
{\bf r}_2(w_1, w_2, w_3)  
& = & {1 \over \sqrt{2}} {1 \over (w + w_1)^{1/2}} 
\left( w_3 \hat{\bf x} + w_2 \hat{\bf z} \right).
\label{r77}
\end{eqnarray}
We again take $\hat{\bf n}(q) = \hat{\bf z}$ and $(\tau_\mu)_k(q) =
\delta_{\mu k}$, and it is easy to see that Eqs.~(\ref{r70}) --
(\ref{r101}) are again satisfied.  Equations~(\ref{r76}) and
(\ref{r77}) exhibit a string singularity on the negative $w_1$-axis,
which consists of shapes satisfying ${\bf r}_1 = 0$.  Intuitively, we
explain the location of the string by the following observation: if
${\bf r}_1 = 0$, then the orientation of ${\bf r}_2$ within the
$xz$-plane is not fixed.  The conventions are valid everywhere off of
the string; there is no branch cut for the $zxz$-frame as there was
for the principal axis frame.  The Taylor series given in
Eq.~(\ref{r11}) is again applicable for the present conventions.  Of course
the coefficients $a_{kn}(w_1, w_2)$ are different here and the domain of
validity is also different.

In the next example, we continue to use the $zxz$-frame, $\hat{\bf
n}(q) = \hat{\bf z}$, and $(\tau_\mu)_k(q) = \delta_{\mu k}$, but use a
different set of hyperspherical coordinates $(\rho, \zeta, \theta)$
defined by

\begin{eqnarray}
w_1 & = & \rho^2 \sin 2 \zeta, \\ 
w_2 & = & \rho^2 \cos 2 \zeta \cos \theta, \\
w_3 & = & \rho^2 \cos 2 \zeta \sin \theta.
\end{eqnarray}
These are the asymmetric hyperspherical coordinates of Smith
\cite{Smith62,Aquilanti93}.  They are compatible everywhere in the physical
region except on the $w_1$-axis, where there is a coordinate
singularity.  Notice that this coordinate singularity includes the
singularities in the $zxz$-frame, which occur on the negative
$w_1$-axis.  The coordinate $\theta$ is transverse to the collinear
shapes, which occur at both $\theta = 0$ and $\theta = \pi$. We
concentrate on the boundary $\theta = 0$ first.  Since the conventions
are valid over the entire half-plane $\theta = 0$, the Taylor series

\begin{equation}
\chi^\ell_k(\rho, \zeta, \theta) 
= \sum_{n = 0}^\infty a_{kn}(\rho,\zeta) \theta^{|k| + 2n} 
\label{r35}
\end{equation}
is sufficient to guarantee analyticity of the external wave function.
With regards to the half-plane $\theta = \pi$, by our definition the
conventions are not valid there because we require the boundary be
given by $q_3 = 0$.  Nevertheless, by using new coordinates $(\rho,
\zeta, \theta' = \pi - \theta)$, the conventions are valid on this
half-plane. If the external wave function is to be analytic on this
half-plane, $\chi^\ell_k$ must satisfy a Taylor series in $\theta'$
identical in form to Eq.~(\ref{r35}) for $\theta$.  One set of
functions which do satisfy the appropriate conditions at both $\theta
= 0$ and $\theta = \pi$ are the associated Legendre polynomials
$P^\ell_k(\theta)$.  An internal wave function $\chi^\ell_k(\rho,
\zeta, \theta) = b_k(\rho,\zeta) P^\ell_k(\theta)$, with $b_k(\rho,
\zeta)$ analytic, therefore lifts to an analytic external wave
function (in the region of validity).  Internal wave functions of this
form arise naturally when constructing hyperspherical harmonics.
(See, for example, Aquilanti et al. \cite{Aquilanti98,Aquilanti86a}.)

\section{The Planar Two-Body Problem}
\label{s3}

Before proving the theorem on three-body boundary conditions, we
discuss the two-body problem in the plane.  Not only is this simpler
case useful practice for the three-body problem, but in fact our proof
of the three-body results relies on the the two-body results presented
here.

First, we must adapt the basic concepts and notation introduced for
the three-body problem for use with the two-body problem.  The
configuration space of a two-body system, in the center of mass frame,
is just the two-dimensional plane, the relative position of one
body with respect to the other being denoted here as ${\bf r}_s
\in {\Bbb R}^2$.  The shape of the two-body system depends only on the
separation distance $r$ between the bodies, and we denote the shape
coordinate by $q(r)$.  We assign to each shape $q$ a reference
orientation ${\bf r}(q)$.  The reference orientation ${\bf r}(q)$ is
simply a point on the circle of radius $r(q)$ centered at the origin
of ${\Bbb R}^2$.  This fact makes the reference orientations much
easier to visualize here than for the three-body problem; the
reference orientations all lie on a curve beginning at the origin and
intersecting each concentric circle once as it moves out to infinity.
An arbitrary configuration ${\bf r}_s$ is given by

\begin{equation}
{\bf r}_s = {\sf R}{\bf r}(q),
\label{r21}
\end{equation}
where ${\sf R} \in SO(2)$ denotes the orientation of the system.  We
will denote the rotation angle of ${\sf R}$ by $\theta$.  

As an example, one could choose the internal coordinate to be $q(r)=r$
and the reference orientation to be

\begin{equation}
{\bf r}(r) = r \hat{\bf x}.
\end{equation}
This choice produces shape and orientation coordinates $(r,
\theta)$ which are the usual polar coordinates on ${\Bbb R}^2$.

We now discuss the two-body wave function $\Psi$. The action of ${\sf
Q} \in SO(2)$ on $\Psi$ is

\begin{equation}
\bbox{(}{\cal R} ( {\sf Q}) \Psi \bbox{)} ({\bf r}_s) 
= \Psi({\sf Q}^T{\bf r}_s).
\label{r37}
\end{equation}
We denote by $\Psi_m$ a wave function which transforms according to
the irrep $m$ of $SO(2)$,

\begin{equation}
\Psi_m({\sf Q}^T{\bf r}_s) 
= \bbox{(} {\cal R}({\sf Q}) \Psi_m \bbox{)} ({\bf r}_s) 
= e^{-i m \alpha} \Psi_m({\bf r}_s),
\label{r1}
\end{equation}
where $\alpha$ is the rotation angle of ${\sf Q}$.  The internal wave
function $\chi_m(q)$ is defined by

\begin{equation}
\chi_m(q) = \Psi_m \bbox{(} {\bf r}(q) \bbox{)}. 
\label{r14}
\end{equation}
Given $\chi_m(q)$ we may recover the external wave function
$\Psi_m({\bf r}_s)$ with the aid of Eqs.~(\ref{r21}), (\ref{r1}), and
(\ref{r14}),

\begin{equation}
\Psi_m({\bf r}_s) 
= \Psi_m \bbox{(}{\sf R} {\bf r}(q) \bbox{)} 
= e^{i m \theta} \chi_m(q).
\label{r13}
\end{equation}
Equations (\ref{r37}) -- (\ref{r13}) are obviously analogous to
Eqs.~(\ref{r36}) -- (\ref{r7}) for the three-body problem.

We take the Cartesian coordinates $(r_{sx}, r_{sy})$ (that is, the
usual space-fixed $(x, y)$) as the privileged coordinates for defining
analyticity of functions on configuration space.  We take the radial
coordinate $r$ as the privileged coordinate for defining analyticity
of functions on shape space.  Since $r$ is a positive quantity, we
must give special consideration to the boundary $r=0$.  A function on
shape space is said to be analytic at $r=0$ if it can be analytically
continued (as a function of $r$) into the region $r<0$.

For the three-body problem, there were four different conventions that
had to be specified.  For the two-body problem, we need only specify
two conventions: the shape coordinate $q$ and the reference
orientation ${\bf r}(q)$; there is no analog of $\hat{\bf n}(q)$ or
$\tau_\mu(q)$ for the two-body problem.  Away from the two-body
collision, these two conventions are said to be ``valid'' if the shape
coordinate $q$ is compatible with $r$ and if ${\bf r}(q)$ is analytic.
At the two-body collision, $q$ must still be compatible with $r$ and
${\bf r}(q)$ must still be analytic.  However, we also require that
$q=0$ coincide with the two-body collision and that the following
conditions be met

\begin{eqnarray}
q(-r) & = & -q(r), \\
{\bf r}(-q) & = & -{\bf r}(q).
\end{eqnarray}

We now state the following two-body theorem, which is analogous to
Theorem \ref{t1} for the three-body problem.

\begin{theorem}
\label{t2}
Let a configuration ${\bf r}_s$ have shape $q$, and assume valid
conventions for the shape coordinate $q$ and the reference orientation
${\bf r}(q)$ in the neighborhood of $q$.

\vskip 12pt
\noindent (i) If $q \ne 0$, then $\Psi_m$ is analytic at ${\bf r}_s$
if and only if $\chi_m$ is analytic at $q$.

\noindent (ii) If $q = 0$, then $\Psi_m$ is analytic at ${\bf r}_s =
0$ if and only if $\chi_m$ is analytic at $0$ with the Taylor series

\begin{equation}
\chi_m (q) = \sum_{n = 0}^\infty a_{mn} q^{|m| + 2n},
\label{r15}
\end{equation}
where the $a_{mn}$ are constant complex coefficients.

\end{theorem}

\noindent \underline{Proof}

For the entirety of this proof, when we say that a function of either
the Cartesian coordinates or shape coordinate is analytic, we mean
only that it is locally analytic at the specific points, ${\bf r}_{s}$
or $q$ respectively, mentioned in the statement of the theorem.

Equation~(\ref{r1}) shows that if $\Psi_m$ is analytic at an arbitrary
${\bf r}_s$, then $\Psi_m$ is analytic at any other orientation ${\sf
Q}^T {\bf r}_s$ with ${\sf Q} \in SO(2)$ arbitrary.  We therefore
assume without loss that the specific configuration ${\bf r}_s$ in the
statement of the Theorem is the reference orientation ${\bf
r}$.

(i) Assume $q \ne 0$.  Assume $\Psi_m({\bf r}_{s \alpha})$ is
analytic.  From Eq.~(\ref{r14}), the fact that ${\bf r}(q)$ is
analytic, and the fact that the composition of analytic functions is
analytic, $\chi_m(q)$ is analytic.

Next assume $\chi_m(q)$ is analytic.  By the assumption of
compatibility, $q(r)$ is analytic.  Furthermore, $r({\bf r}_s) =
(r_{sx}^2 + r_{sy}^2)^{1/2}$ is an analytic function of the Cartesian
coordinates.  Hence, $q({\bf r}_s) = q \bbox{(} r({\bf r}_s) \bbox{)}$
is analytic.  Turning to the rotation matrix ${\sf R}$ in
Eq.~(\ref{r21}), its rotation angle $\theta$ is given by $\theta({\bf
r}_s, {\bf r}) = \arcsin \bbox{(} \hat{\bf z} \cdot ({\bf r}
\bbox{\times} {\bf r}_s /r^2)\bbox{)}$, which is an analytic function
of ${\bf r}_s$ and ${\bf r}$.  (We only consider $\theta$ in the range
$-\pi/2 < \theta <
\pi/2$ since we are only checking for analyticity at the
reference orientation $\theta = 0$.)  Furthermore, ${\bf r}({\bf r}_s)
= {\bf r}\bbox{(}q({\bf r_s})\bbox{)}$ is analytic since ${\bf r}(q)$ is analytic by
the validity assumption and $q({\bf r}_s)$ was shown to be analytic
above.  Hence,

\begin{equation}
\theta({\bf r}_s) = \theta \bbox{(} {\bf r}_s, {\bf r}({\bf r}_s) \bbox{)}
\end{equation}
is analytic.  Since $\exp(im\theta)$ is an analytic function of
$\theta$, we conclude that

\begin{equation}
\Psi_m({\bf r}_s) = e^{i m \theta({\bf r}_s)} \chi_m \bbox{(} q({\bf r}_s) \bbox{)}
\end{equation}
is an analytic function of the Cartesian coordinates.

(ii) Assume $q=0$.  Assume $\Psi_m({\bf r}_{s \alpha})$ is analytic.
By the same argument as in case (i), $\chi_m(q)$ is analytic at $q=0$.
To prove Eq.~(\ref{r15}), we differentiate Eq.~(\ref{r1}) $d$ times
with respect to ${\bf r}_s$,

\begin{equation}
\sum_{k_1 ... k_d} Q_{j_1 k_1} ... Q_{j_d k_d} 
\left( {\partial \over \partial r_{s k_1}} \cdots {\partial \over \partial r_{s k_d}} 
\Psi_m \right) ({\sf Q}^T {\bf r}_s) 
= e^{- i m \alpha} 
\left( {\partial \over \partial r_{s j_1}} \cdots {\partial \over \partial r_{s j_d}} 
\Psi_m\right)  ({\bf r}_s),
\label{r2}
\end{equation}
where $Q_{j k}$, $j,k = x,y$, are the components of ${\sf Q}$.
Evaluating Eq.~(\ref{r2}) at ${\bf r}_s = 0$ produces

\begin{equation}
\sum_{k_1 ... k_d} Q_{j_1 k_1} ... Q_{j_d k_d} 
\left( {\partial \over \partial r_{s k_1}} \cdots {\partial \over \partial r_{s k_d}} 
\Psi_m \right) ( 0 ) 
= e^{- i m \alpha} 
\left( {\partial \over \partial r_{s j_1}} \cdots {\partial \over \partial r_{s j_d}} 
\Psi_m \right)( 0 ).
\label{r3}
\end{equation}
Equation~(\ref{r3}) shows that the rank $d$ tensor $(\partial /
\partial r_{s j_1} ... \partial / \partial r_{s j_d} \Psi_m) ( 0 )$
transforms under the completely symmetrized action of $SO(2)$ as an
irrep of $SO(2)$ labeled by $m$.  (See Appendix.)  The decomposition
of the fully symmetrized action of $SO(2)$ on rank $d$ tensors
decomposes into irreps as shown in Eq.~(\ref{r4}).  If $m$ does not
label one of the irreps included in
Eq.~(\ref{r4}), that is, if $m \ne d, d-2, ..., -(d-2), -d$, then
$(\partial / \partial r_{s j_1} ... \partial / \partial r_{s j_d}
\Psi_m) ( 0 ) = 0$. Consequently, the chain rule and Eq.~(\ref{r14})
show that

\begin{equation}
\left({d^d\over d q^d} \chi_m\right) (0) 
= \sum_{k_1 ... k_d} {d r_{s k_1} \over dq}(0) \cdots {d r_{s k_d}
\over dq}(0)
\left( {\partial \over \partial r_{sk_1}} \cdots {\partial \over
\partial r_{sk_d}}\Psi_m \right)( 0 ) = 0
\end{equation}
when $m \ne d, d-2, ..., -(d-2), -d$.  This shows that the appropriate Taylor
coefficients vanish in order to produce the Taylor series in
Eq.~(\ref{r15}).

Now assume that $\chi_m$ is analytic at $q=0$ with Taylor series shown
in Eq.~(\ref{r15}).  We first complete the proof of Theorem~\ref{t2}
assuming the internal coordinate $q(r)=r$ and the reference
configuration ${\bf r}(r) = r \hat{\bf x}$.  We explicitly construct a
function $\tilde{\Psi}_m({\bf r}_s)$ by

\begin{equation}
\tilde{\Psi}_m({\bf r}_s) 
= \sum_{n = 0}^\infty \sum_{k_1 ... k_{|m | + 2n}}
a_{mn} \left(t^{|m| + 2n}_m\right)_{k_1 ... k_{|m| + 2n}} 
r_{s k_1} ... r_{s k_{|m| + 2n}},
\label{r56}
\end{equation}
where the $a_{mn}$ are the same Taylor coefficients as in
Eq.~(\ref{r15}) and the $t^{|m| + 2n}_m$ are the rank $|m| + 2n$
tensors defined by Eq.~(\ref{r17}).  Clearly the transformation
property Eq.~(\ref{r16}) of the $t^{|m| + 2n}_m$ shows that
$\tilde{\Psi}_m$ satisfies Eq.~(\ref{r1}).  Hence, we may apply the
same analysis to $\tilde{\Psi}_m$ as we have for $\Psi_m$.  In
particular, $\tilde{\Psi}_m$ is uniquely determined via Eq.~(\ref{r13})
by the internal wave function $\tilde{\chi}_m$ defined by Eq.~(\ref{r14})

\begin{equation}
\tilde{\chi}_m(r) = \tilde{\Psi}_m(r \hat{\bf x})
=  \sum_{n = 0}^\infty a_{mn} r^{|m| + 2n},
\label{r50}
\end{equation}
where we have used Eqs.~(\ref{r56}) and (\ref{r38}).  Since $\chi_m$ and
$\tilde{\chi}_m$ have the same Taylor series, $\chi_m =
\tilde{\chi}_m$.  Furthermore, the unique correspondence between
internal and external wave functions guarantees that $\Psi_m =
\tilde{\Psi}_m$.  From Eq.~(\ref{r56}), we see that $\Psi_m =
\tilde{\Psi}_m$ is analytic at $0$ by construction.  This completes
the proof of Theorem~\ref{t2} for the specific conventions chosen
above.

We mention two noteworthy special cases of Theorem \ref{t2}, which the
above analysis has now proven.  First, if $f(r)$ is an even analytic
function, then $f({\bf r}_s) = f \bbox{(} r({\bf r}_s) \bbox{)}$ is an
analytic function of the Cartesian coordinates.  (This fact is quite
trivial to prove from scratch by simply considering the Taylor series
of the two functions.)  Second, if $f(r)$ is an odd analytic function,
then $f({\bf r}_s) = f(r, \theta) =
\exp(i\theta) f(r)$ is an analytic function of the Cartesian
coordinates, where $(r, \theta)$ are the standard polar coordinates.

We now assume arbitrary valid conventions ${\bf r}(q)$ and $q(r)$ for
the reference orientation and shape coordinate respectively.  To
complete the proof of Theorem~\ref{t2} for these conventions, we first
define and discuss three useful functions ${\sf F}({\bf r}_s)$, ${\bf G}({\bf
r}_s)$, and ${\bf H}({\bf r}_s)$ related to these conventions.  By the
validity assumptions, ${\bf r}(r)$ is an odd analytic function in the
neighborhood of $r=0$, and hence $\hat{\bf r}(r) = {\bf r}(r)/r$ is an
even analytic function in the neighborhood of $r=0$.  Define the
matrix-valued function ${\sf F}(r)$ by

\begin{equation}
{\sf F}(r) 
= \hat{\bf x} \hat{\bf r}^T(r) 
+ \hat{\bf y} \bbox{(} \hat{\bf z} \bbox{\times} \hat{\bf r}(r) \bbox{)}^T.
\label{r51}
\end{equation}
Notice that ${\sf F}(r)$ is the unique matrix in $SO(2)$ satisfying

\begin{equation}
{\sf F}(r) \hat{\bf r}(r) = \hat{\bf x}.
\label{r41}
\end{equation}
Since $\hat{\bf r}(r)$ is an even analytic function, we see from
Eq.~(\ref{r51}) that ${\sf F}(r)$ is an even analytic function.  Hence, from
the first of the two special cases of Theorem~\ref{t2} mentioned
above, ${\sf F}({\bf r}_s) = {\sf F} \bbox{(} r({\bf r}_s) \bbox{)}$
is also analytic.  Notice of course that ${\sf F}({\bf
r}_s)$ is invariant under all rotations ${\sf Q} \in SO(2)$,

\begin{equation}
{\sf F}({\sf Q}{\bf r}_s) = {\sf F}({\bf r}_s).
\label{r59}
\end{equation}

By the validity assumptions, $q(r)$ is an odd analytic function of
$r$.  Hence, from the second of the two special cases of
Theorem~\ref{t2} mentioned above, the function

\begin{equation}  
G({\bf r}_s) = e^{i \theta} q(r)
\end{equation}
is an analytic function of the Cartesian coordinates.  (Here $(\theta,
r)$ are the standard polar coordinates in ${\Bbb R}^2$.)  We define a
vector-valued version of $G({\bf r}_s)$, denoted ${\bf G}({\bf r}_s)$, by

\begin{equation}
{\bf G}({\bf r}_s) 
= \text{Re} \bbox{(} G({\bf r}_s) \bbox{)} \hat{\bf x} +
\text{Im} \bbox{(} G({\bf r}_s) \bbox{)} \hat{\bf y}
= q(r) \hat{\bf r}_s.
\label{r39}
\end{equation}
Obviously, ${\bf G}({\bf r}_s)$ is also an analytic function of the
Cartesian coordinates.  Recall that the compatibility of the
coordinate $q$ guarantees that $q(r)$ has an inverse, which we denote
by $q^{-1}$, that is, $q^{-1} \bbox{(} q(r) \bbox{)} = r$.  The function ${\bf
G}({\bf r}_s)$ therefore has an inverse given by

\begin{equation}
{\bf G}^{-1}({\bf r}_s) = q^{-1}(r)\hat{\bf r}_s,
\label{r55}
\end{equation} 
as may be verified by inserting this formula into Eq.~(\ref{r39}).
Equations~(\ref{r39}) and (\ref{r55}) easily admit the following
results

\begin{eqnarray}
{\bf G}({\sf Q}{\bf r}_s) & = & {\sf Q} {\bf G}({\bf r}_s), 
\label{r57} \\
{\bf G}^{-1}({\sf Q}{\bf r}_s) & = & {\sf Q} {\bf G}^{-1}({\bf r}_s),
\label{r58}
\end{eqnarray}
where ${\sf Q} \in SO(2)$ is arbitrary.

We define a new function ${\bf H}({\bf r}_s)$ by

\begin{equation}
{\bf H}({\bf r}_s) = {\sf F}({\bf r}_s){\bf G}({\bf r}_s).
\label{r52}
\end{equation}
This function has several important properties.  First, since both
${\bf G}({\bf r}_s)$ and ${\sf F}({\bf r}_s)$ are analytic, ${\bf
H}({\bf r}_s)$ is analytic.   Second,

\begin{equation}
{\bf H} \bbox{(} {\bf r}(q) \bbox{)}
= {\sf F} \bbox{(} {\bf r}(q) \bbox{)} {\bf G} \bbox{(} {\bf r}(q) \bbox{)}
= q\hat{\bf x},
\label{r62}
\end{equation}
which follows from Eqs.~(\ref{r41}) and (\ref{r39}).  A third fact is
that ${\bf H}$ is invertible.  This fact requires more work to prove,
which we do by explicitly constructing ${\bf H}^{-1}$.  Let ${\bf
r}_s' = {\bf H}({\bf r}_s)$.  Then,

\begin{equation}
{\bf r}_s 
= {\bf G}^{-1} \bbox{(} {\sf F}^{-1}({\bf r}_s) {\bf r}_s'\bbox{)} 
= {\sf F}^{-1}({\bf r}_s){\bf G}^{-1}({\bf r}_s'),
\label{r54}
\end{equation}
where we use the definition of ${\bf H}$ and the fact that ${\bf G}$
is invertible in the first equality and the second equality follows
from Eq.~(\ref{r58}).  Now, the magnitudes of ${\bf r}_s'$ and ${\bf
r}_s$ are related by

\begin{equation}
| {\bf r}_s'| 
= | {\sf F}({\bf r}_s) {\bf G}({\bf r}_s) | 
= | {\bf G}({\bf r}_s) | 
= q(|{\bf r}_s|),
\end{equation}
where we have used Eq.~(\ref{r39}) in the final equality.  Turning this
relation around and using Eq.~(\ref{r55}) we have,

\begin{equation}
|{\bf r}_s| 
= q^{-1}(|{\bf r}_s'|) 
= |{\bf G}^{-1}({\bf r}_s')|.
\label{r53}
\end{equation}
As witnessed by Eq.~(\ref{r59}), ${\sf F}({\bf r}_s)$ depends only on
the magnitude of its argument, and hence Eqs.~(\ref{r54}) and (\ref{r53})
combine to produce

\begin{equation}
{\bf r}_s 
= {\sf F}^{-1} \bbox{(} {\bf G}^{-1}({\bf r}_s') \bbox{)} {\bf G}^{-1}({\bf r}_s'),
\end{equation}
which gives ${\bf r}_s$ in terms of ${\bf r}'_s$.  Hence,

\begin{equation}
{\bf H}^{-1}({\bf r}_s') 
= {\sf F}^{-1} \bbox{(} {\bf G}^{-1}({\bf r}_s') \bbox{)} {\bf G}^{-1}({\bf r}_s').
\end{equation}
Another fact regarding ${\bf H}$ is that if ${\sf Q} \in SO(2)$ is
arbitrary, then

\begin{equation}
{\bf H}({\sf Q}{\bf r}_s) 
= {\sf F}({\sf Q}{\bf r}_s) {\bf G} ({\sf Q}{\bf r}_s)
= {\sf F}({\bf r}_s) {\sf Q} {\bf G} ({\bf r}_s)
= {\sf Q} {\sf F}({\bf r}_s) {\bf G} ({\bf r}_s)
= {\sf Q} {\bf H}({\bf r}_s),
\label{r40}
\end{equation}
where the second equality follows from Eqs.~(\ref{r59}) and
(\ref{r57}) and the third equality follows from the commutativity of
the group $SO(2)$. From Eq.~(\ref{r40}) follows an analogous identity
for ${\bf H}^{-1}$

\begin{equation}
{\bf H}^{-1}({\sf Q}{\bf r}_s) = {\sf Q}{\bf H}^{-1}({\bf r}_s).
\label{r60}
\end{equation}

We now have the proper background to complete the proof.  Since ${\bf
H}$ is invertible, we introduce the function $\tilde{\Psi}_m$ by

\begin{eqnarray}
\tilde{\Psi}_m({\bf r}_s) & = & \Psi_m \bbox{(} {\bf H}^{-1}({\bf r}_s) \bbox{)},
\label{r61} \\
\Psi_m({\bf r}_s) & = & \tilde{\Psi}_m \bbox{(} {\bf H}({\bf r}_s) \bbox{)}.
\label{r63}
\end{eqnarray}
We see from Eq.~(\ref{r60}) that $\tilde{\Psi}_m$ satisfies
Eq.~(\ref{r1}) since $\Psi_m$ satisfies Eq.~(\ref{r1}).  We define the
internal wave function $\tilde{\chi}_m(q)$ using the old
convention ${\bf r}(q) = q\hat{\bf x}$,

\begin{equation}
\tilde{\chi}_m(q) = \tilde{\Psi}_m(q\hat{\bf x}).
\end{equation}
Using the new convention ${\bf r}(q)$, we have $\chi_m(q)$ given by

\begin{equation}
\chi_m(q) = \Psi_m \bbox{(} {\bf r}(q) \bbox{)}
= \tilde{\Psi}_m({\bf H} \bbox{(} {\bf r}(q) \bbox{)} )
= \tilde{\Psi}_m(q\hat{\bf x}),
\end{equation}
where the second equality follows from Eq.~(\ref{r63}) and the third
from Eq.~(\ref{r62}).  Thus, the functional form of
${\chi}_m(q)$ and $\tilde{\chi}_m(q)$ are identical.  Since we
have assumed that $\chi_m$ is analytic at $q=0$ with the Taylor series
in Eq.~(\ref{r15}), $\tilde{\chi}_m$ is also analytic with the
identical Taylor series.  Now since we have proved Theorem~\ref{t2}
for the old conventions used to define $\tilde{\chi}_m(q)$, we have
that $\tilde{\Psi}_m({\bf r}_s)$ is an analytic function of the
Cartesian coordinates.  Furthermore, since ${\bf H}({\bf r}_s)$ is
analytic, Eq.~(\ref{r63}) implies that $\Psi_m({\bf r}_s)$ is
analytic. ${\cal QED}$.

\section{Proof of Three-Body Boundary Conditions}
\label{s4}

In this section, we prove Theorem \ref{t1}.

\noindent \underline{Proof}

For the entirety of this proof, when we say that a function of either
the Jacobi coordinates or shape coordinates is analytic, we mean only
that it is locally analytic at the specific points, ${\bf r}_{s
\alpha}$ or $q$ respectively, mentioned in the theorem.

A common fact we will use several times is that the Wigner matrices
${\cal D}^\ell_{mk}({\sf Q})$ are analytic functions of the rotation
matrices ${\sf Q} \in SO(3)$.  Equation~(\ref{r5}) thus shows that if
$\Psi^\ell_m$ is analytic at an arbitrary ${\bf r}_{s \alpha}$, then
$\Psi^\ell_m$ is analytic at any other orientation ${\sf Q}^T {\bf
r}_{s\alpha}$ with ${\sf Q}
\in SO(3)$ arbitrary.  We therefore assume without loss that the
specific configuration ${\bf r}_{s\alpha}$ in the statement of the
Theorem is the reference orientation ${\bf r}_\alpha$.

(i) Assume $q$ is noncollinear.  The proof here is a straightforward
generalization of the proof of part (i) of Theorem~\ref{t2}.  First
assume $\Psi^\ell_m({\bf r}_{s \alpha})$ is analytic. From
Eqs.~(\ref{r6}) and (\ref{r46}), the fact that both ${\bf
r}_\alpha(q)$ and $(\tau_\mu)_k(q)$ are analytic, and the fact that
the composition of analytic functions is analytic, $\chi^\ell_\mu(q)$
and $\chi^\ell_k(q)$ are both analytic.

Next assume $\chi^\ell_\mu(q)$ is analytic.  From Eq.~(\ref{r79}) and
the fact that $(\tau_\mu)_k(q)$ is analytic, $\chi^\ell_k(q)$ is also
analytic.  The validity assumption guarantees that $q(w_\mu)$ is
analytic.  From Eqs.~(\ref{r32}) -- (\ref{r34}) it is evident that the
functions $w_\mu({\bf r}_{s \alpha})$ are themselves analytic.
Therefore, $q({\bf r}_{s \alpha}) = q \bbox{(} w_\mu({\bf r}_{s
\alpha}) \bbox{)}$ is analytic.

We now focus on the orientation matrix ${\sf R}$ in Eq.~(\ref{r20})
which may be  expressed in terms of the vectors ${\bf r}_{s
\alpha}$ and ${\bf r}_\alpha$ as

\begin{eqnarray}
{\sf R}({\bf r}_{s \alpha}, {\bf r}_{\alpha}) 
& = & {1 \over |{\bf r}_1 \bbox{\times} {\bf r}_2|^2} 
\left[ {\bf r}_{ s 1} {\bf v}_1^T + {\bf r}_{s2} {\bf v}_2^T
+ ({\bf r}_{s1} \bbox{\times} {\bf r}_{s2})
({\bf r}_1 \bbox{\times} {\bf r}_2)^T \right]
\label{r10}, \\
{\bf v}_1 & = & -{\bf r}_2 \bbox{\times}({\bf r}_2 \bbox{\times} {\bf r}_1), 
\label{r99} \\
{\bf v}_2 & = & -{\bf r}_1 \bbox{\times}({\bf r}_1 \bbox{\times} {\bf r}_2).
\label{r100}
\end{eqnarray}
To confirm the above expression for ${\sf R}$, we need only verify
Eq.~(\ref{r20}) and show that ${\sf R}({\bf r}_1 \bbox{\times}{\bf
r}_2) ={\bf r}_{s1} \bbox{\times}{\bf r}_{s2}$, both of which follow
easily from the simple identities

\begin{eqnarray}
{\bf v}_\alpha \cdot {\bf r}_\alpha 
& = & |{\bf r}_1 \bbox{\times} {\bf r}_2 |^2, \label{r97}\\
{\bf v}_\alpha \cdot ({\bf r}_1 \bbox{\times} {\bf r}_2) 
& = & 0, \\
{\bf v}_1 \cdot {\bf r}_2 = {\bf v}_2 \cdot {\bf r}_1
& = & 0.
\end{eqnarray} 
Since ${\bf r}_1 \bbox{\times} {\bf r}_2$ does not vanish, it is clear
that ${\sf R}({\bf r}_{s \alpha}, {\bf r}_\alpha)$ is an analytic
function of the vectors ${\bf r}_{s
\alpha}$ and ${\bf r}_\alpha$. Since both ${\bf
r}_\alpha(q)$ and $q({\bf r}_{s \alpha})$ are analytic, the function
${\bf r}_\alpha({\bf r}_{s \alpha}) = {\bf r}_\alpha\bbox{(} q({\bf r}_{s
\alpha}) \bbox{)}$ is analytic.  Hence, ${\sf R}({\bf r}_{s \alpha}) 
= {\sf R} \bbox{(} {\bf r}_{s \alpha}, {\bf r}_{\alpha}({\bf r}_{s
\alpha}) \bbox{)}$ is also analytic.  Recalling that the Wigner
matrices are analytic functions of ${\sf R}$ and that $q({\bf r}_{s
\alpha})$ is analytic, we see that

\begin{equation}
\Psi^\ell_m({\bf r}_{s\alpha}) 
= \sum_k {\cal D}^{\ell *}_{ m k} \bbox{(} {\sf R}({\bf r}_{s \alpha}) \bbox{)}
\chi^\ell_k \bbox{(} q({\bf r}_{s \alpha}) \bbox{)}
\end{equation}
is an analytic function of the Jacobi coordinates.

(ii) Assume $q$ is collinear.  We define a new set of coordinates on
configuration space consisting of two shape coordinates, two
orientation coordinates, and two coordinates constructed from the one
remaining shape coordinate and the one remaining orientation
coordinate.  We call these coordinates the ``mixed coordinates''.  The
majority of the remainder of the proof will be dedicated to defining
the mixed coordinates and proving the most important fact about them,
that they are compatible with the Jacobi coordinates at the collinear
configuration.

First, we discuss the parameterization of the rotation matrix ${\sf
R}$ in Eq.~(\ref{r20}).  Consider an arbitrary unit vector $\hat{\bf
e}$ lying in the northern hemisphere (excluding
the equator), with $\hat{\bf n}(q)$ as the north pole.  We define
${\sf U}(q,\hat{\bf e})
\in SO(3)$ to be the unique rotation such that

\begin{equation}
{\sf U}(q,\hat{\bf e}) \hat{\bf n}(q)
= \hat{\bf e}
\label{r67}
\end{equation}
and such that its rotation axis lies on the equator.  Specifically,
the rotation axis of ${\sf U}$ lies in the direction of

\begin{equation}
{\bf a}(q, \hat{\bf e}) = \hat{\bf n}(q) \bbox{\times} \hat{\bf e},
\label{r65}
\end{equation}  
and the rotation angle of ${\sf U}$ has the value $\arcsin |{\bf a}|$.
Thus, ${\sf U}(q, \hat{\bf e})$ is explicitly given by
	
\begin{eqnarray}
{\sf U}(q, \hat{\bf e}) 
& = & \exp[f(|{\bf a}(q, \hat{\bf e})|)  \bbox{(} {\bf a}(q, \hat{\bf e}) \bbox{\times} \bbox{)} ], 
\label{r83} \\
f(x)  & =  & {\arcsin x \over x},
\label{r93}
\end{eqnarray}
where we have introduced the notation ${\bf a} \bbox{\times}$ for the
$3\times3$ matrix which maps an arbitrary vector ${\bf v}$ into ${\bf
a} \bbox{\times} {\bf v}$.  The purpose of Eqs.~(\ref{r83}) and
(\ref{r93}) is to demonstrate the analyticity of ${\sf U}(q, \hat{\bf
e})$.  Observe that $\arcsin(x)$ is an odd analytic function on the
interval $(-1, 1)$, and hence, $f(x)$ is an even analytic function on
$(-1,1)$.  This in turn implies that $f({\bf a}) = f(|{\bf a}|)$ is an
analytic function of ${\bf a}$.  From Eq.~(\ref{r65}) and the fact
that $\hat{\bf n}(q)$ is analytic at $q_3 = 0$, ${\bf a}(q,
\hat{\bf e})$ is analytic at $q_3 = 0$ and at all $\hat{\bf e}$ in the
northern hemisphere defined by the north pole $\hat{\bf n}(q_3 = 0)$.
The analyticity of the exponential function permits the following
final statement.  The function ${\sf U}(q, \hat{\bf e})$ expressed in
Eq.~(\ref{r83}) is an analytic function at $q_3 = 0$ and at all
$\hat{\bf e}$ in the northern hemisphere with $\hat{\bf n}(q_3 = 0)$
as the north pole.

For an arbitrary rotation angle $\theta$ we define ${\sf V}(q,
\theta)$ to be the rotation by $\theta$ about $\hat{\bf n}(q)$.
Explicitly, this rotation is given by

\begin{equation}
{\sf V}(q, \theta)
= \hat{\bf n}(q) \hat{\bf n}^T(q) 
+ [\sin \theta \bbox{(} \hat{\bf n}(q)\bbox{\times} \bbox{)} + \cos \theta]{\sf P}_\perp(q).
\label{r64}
\end{equation}
We now express the rotation matrix
${\sf R}$ in Eq.~(\ref{r20}) by

\begin{equation}
{\sf R}(\hat{\bf e}, \theta) 
= {\sf U}(q,\hat{\bf e}) {\sf V}(q, \theta),
\label{r49}
\end{equation}
where we have parameterized ${\sf R}$ by the quantities $\hat{\bf e}$
and $\theta$.

We assume without loss of generality that $\hat{\bf n}(q_3 = 0) =
\hat{\bf z}$.  (This assumption simply amounts to a judicious choice
of space frame.)  We may thus define ${\sf W}(q) \in SO(3)$ in the
neighborhood of $q_3 = 0$ to be the unique matrix such that

\begin{equation}
{\sf W}(q) \hat{\bf z} = \hat{\bf n}(q)
\label{r43}
\end{equation}
and such that its axis of rotation lies in the $xy$-plane.  In fact,
we see from Eq.~(\ref{r67}) that 

\begin{equation}
{\sf W}(q) = {\sf U}^T(q, \hat{\bf z}).
\label{r84}
\end{equation}
The analyticity property of ${\sf U}$ implies that ${\sf W}(q)$ is
analytic at $q_3 = 0$.  We use ${\sf W}(q)$ to define an orthonormal
frame $\hat{\bf n}_i(q), i =1,2,3$, by

\begin{eqnarray}
\hat{\bf n}_1(q) & = & {\sf W}(q) \hat{\bf x}, 
\label{r85} \\
\hat{\bf n}_2(q) & = & {\sf W}(q) \hat{\bf y}, 
\label{r86} \\
\hat{\bf n}_3(q) & = & {\sf W}(q) \hat{\bf z} = \hat{\bf n}(q). 
\label{r87} 
\end{eqnarray}
The functions $\hat{\bf n}_i(q)$ are of course analytic.  

The unit vector $\hat{\bf e}$ is determined by only
two of its components, of which we choose $\hat{e}_1 = \hat{\bf e}
\cdot \hat{\bf n}_1$ and $\hat{e}_2 = \hat{\bf e} \cdot \hat{\bf
n}_2$.  The third component $\hat{e}_3(\hat{e}_1, \hat{e}_2) =
(\hat{e}_1^2 +
\hat{e}_2^2)^{1/2}$ is an analytic function of the other two in the
northern hemisphere.  The function 

\begin{equation}
\hat{\bf
e}(q,\hat{e}_1, \hat{e}_2) 
= \hat{e}_1 \hat{\bf n}_1(q) 
+ \hat{e}_2 \hat{\bf n}_2(q)
+ \hat{e}_3(\hat{e}_1, \hat{e}_2) \hat{\bf n}_3(q)
\label{102}
\end{equation}
is of course analytic. 

We next define a pair of vectors ${\bf s}_\alpha (q,\theta)$, $\alpha
= 1,2$, by

\begin{equation}
{\bf s}_\alpha (q,\theta) 
= {\sf V}(q,\theta) {\bf r}_\alpha(q)
= (\hat{\bf n}\cdot{\bf r}_\alpha)(q) \hat{\bf n}(q)
+ \left[ \sin \theta \bbox{(} \hat{\bf n}(q)\bbox{\times} \bbox{)} 
+ \cos\theta\right]({\sf P}_\perp {\bf r}_\alpha)(q),
\label{r66}
\end{equation}
where we have used Eq.~(\ref{r64}).  We introduce two new variables
$u_1$ and $u_2$ by

\begin{eqnarray}
u_1 & = & q_3 \cos \theta, 
\label{r23}\\
u_2 & = & q_3 \sin \theta. 
\label{r26}
\end{eqnarray}
We define $u = q_3$, which is convenient notation since when $u$ is
positive, it is the radial coordinate in $u_1 u_2$-space, that is, $u
= (u_1^2 + u_2^2)^{1/2}$.  The vectors ${\bf s}_\alpha$ are
conveniently parameterized by the new coordinates

\begin{equation}
{\bf s}_\alpha(u_1, u_2)
= (\hat{\bf n}\cdot{\bf r}_\alpha)(u) \hat{\bf n}(u)
+ \left[ u_2 \bbox{(} \hat{\bf n}(u)\bbox{\times} \bbox{)} 
+ u_1 \right]\left[ {({\sf P}_\perp {\bf r}_\alpha)(u) \over u} \right],
\end{equation}
where we have dropped the explicit dependence on $q_1$ and $q_2$.
Since $f(u)$ is an even analytic function of $u$, $f(u_1, u_2) =
f\bbox{(} (u_1^2 + u_2^2)^{1/2} \bbox{)}$ is analytic in $u_1$ and $u_2$.
This fact, together with Eqs.~(\ref{r72}) -- (\ref{r74}), shows that
$\hat{\bf n}$, $\hat{\bf n}\cdot{\bf r}_\alpha$, $({\sf P}_\perp {\bf
r}_\alpha)/u$, and hence ${\bf s}_\alpha$ are all analytic functions
of $u_1$ and $u_2$.  Reintroducing the explicit dependence on $q_1$
and $q_2$, we see that ${\bf s}_\alpha(q_1, q_2, u_1, u_2)$ is
analytic.

Using similar arguments as above, from Eqs.~(\ref{r65}) and
(\ref{r83}) we see that both ${\bf a}(q_1, q_2, u_1, u_2, \hat{\bf e}
= \hat{\bf z})$ and ${\sf U}(q_1, q_2, u_1, u_2, \hat{\bf e} =
\hat{\bf z})$ are analytic.  Hence, Eq.~(\ref{r84}) shows that 
${\sf W}(q_1, q_2, u_1, u_2)$ is analytic, from which follows, using
Eqs.~(\ref{r85}) -- (\ref{r87}), that the $\hat{\bf n}_i(q_1, q_2,
u_1, u_2)$, $i=1,2,3$, are analytic. Hence, Eq.~(\ref{102}) shows that
$\hat{\bf e}(q_1, q_2, u_1, u_2, \hat{e}_1,
\hat{e}_2)$ is analytic.  From this result we also have, using
Eqs.~(\ref{r65}) and (\ref{r83}), that 
${\bf a}(q_1, q_2, u_1, u_2, \hat{e}_1, \hat{e}_2) 
= {\bf a} \bbox{(} q_1, q_2, u_1, u_2, 
\hat{\bf e}(q_1, q_2, u_1, u_2, \hat{e}_1,\hat{e}_2) \bbox{)}$
and
${\sf U}(q_1, q_2, u_1, u_2, \hat{e}_1, \hat{e}_2) 
= {\sf U}\bbox{(}q_1, q_2, u_1, u_2, 
\hat{\bf e}(q_1, q_2, u_1, u_2, \hat{e}_1,\hat{e}_2) \bbox{)}$
are both analytic.

We define the mixed coordinates to be the variables $(q_1, q_2, u_1,
u_2, \hat{e}_1, \hat{e}_2)$.  We can express ${\bf r}_{s \alpha}$ in
terms of the mixed coordinates as follows

\begin{equation}
{\bf r}_{s \alpha} (q_1, q_2, u_1, u_2, \hat{e}_1, \hat{e}_2) 
= {\sf R} {\bf r}_\alpha 
= {\sf U} {\sf V} {\bf r}_\alpha  
=  {\sf U}( q_1, q_2, u_1, u_2, \hat{e}_1, \hat{e}_2) 
{\bf s}_\alpha(q_1, q_2, u_1, u_2).
\end{equation}
where we have used Eqs.~(\ref{r20}), (\ref{r49}), and (\ref{r66}).
Since we have already shown that both ${\sf U}( q_1, q_2, u_1, u_2,
\hat{e}_1, \hat{e}_2)$ and ${\bf s}_\alpha(q_1, q_2, u_1, u_2)$ are
analytic, we see that the Jacobi coordinates are analytic functions of
the mixed coordinates.

To show compatibility of the mixed coordinates with the 
Jacobi coordinates we need now only show that the mixed coordinates
are analytic functions of the Jacobi coordinates.  First,
since the internal coordinates $q_\mu$ are valid coordinates, they are
analytic functions of the $w$-coordinates.  By inspecting
Eqs.~(\ref{r32}) and (\ref{r33}), we see that $w_1({\bf r}_{s
\alpha})$ and $w_2({\bf r}_{s \alpha})$ are analytic, even at a
collinear shape.  However, because of the absolute value in
Eq.~(\ref{r34}) the same can not be said of $w_3({\bf r}_{s \alpha})$.
However, the function $w_3^2({\bf r}_{s \alpha})$ is analytic.  Thus,
we have the following lemma: Any analytic function of $(w_1, w_2,
w_3^2)$ may be viewed, via composition, as an analytic function of
${\bf r}_{s \alpha}$.  Using this lemma and noting that $q_1(w_\mu)$
and $q_2(w_\mu)$ are analytic functions which by Eq.~(\ref{r70}) are
also even in $w_3$, we see that the two mixed coordinates $q_1({\bf
r}_{s \alpha})$ and $q_2({\bf r}_{s \alpha})$ are both analytic
functions of the Jacobi coordinates.

Note that the coordinate $q_3({\bf r}_{s \alpha})$ is not analytic.
However, since $q_3^2(w_\mu)$ is analytic and by Eq.~(\ref{r71}) also
even in $w_3$, $q_3^2({\bf r}_{s \alpha})$ is analytic.  This fact
allows us to generalize our previous lemma regarding the
$w$-coordinates to arbitrary $q$.  As we will have frequent need of
this more general lemma, we record it below.

\begin{lemma}
\label{l1}
If $f(q)$ is an analytic function which is even in $q_3$, then $f({\bf
r}_{s \alpha}) = f \bbox{(} q({\bf r}_{s \alpha}) \bbox{)}$ is
analytic.
\end{lemma}
This lemma, together with Eqs.~(\ref{r72}) -- (\ref{r74}), proves the
analyticity of the following functions: $({\bf r}_\alpha \cdot
\hat{\bf n})({\bf r}_{s
\alpha})$, $|{\sf P}_\perp{\bf r}_\alpha|^2({\bf r}_{s \alpha})$,
$r_\alpha^2({\bf r}_{s \alpha})$, $\hat{\bf n}({\bf r}_{s \alpha})$.
Furthermore, Eqs.~(\ref{r65}) and (\ref{r83}) show that ${\sf U}$ is
even in $q_3$ and hence ${\sf U}({\bf r}_{s \alpha}, \hat{\bf e}) =
{\sf U} \bbox{(} q({\bf r}_{s \alpha}), \hat{\bf e} \bbox{)}$ is
analytic at the collinear configuration and at all $\hat{\bf e}$ in
the northern hemisphere with $\hat{\bf n}(q_3 = 0)$ at the north pole.
Equation~(\ref{r84}) proves the analyticity of ${\sf W}({\bf r}_{s
\alpha}) = {\sf W} \bbox{(} q({\bf r}_{s \alpha}) \bbox{)}$ and hence
Eqs.~(\ref{r85}) -- (\ref{r87}) prove the analyticity of the basis
vectors $\hat{\bf n}_i({\bf r}_{s \alpha}) = \hat{\bf n}_i \bbox{(}
q({\bf r}_{s \alpha}) \bbox{)}$, $i=1,2,3.$

The vector $\hat{\bf e}$ is determined by the equation

\begin{equation}
\hat{\bf e} = {\sf R} \hat{\bf n},
\end{equation}
which follows from Eqs.~(\ref{r64}), (\ref{r49}) and (\ref{r67}).  We
observed earlier that away from a collinear shape, the matrix ${\sf
R}$ is given by Eq.~(\ref{r10}), which results in the following
formula for $\hat{\bf e}$

\begin{eqnarray}
\hat{\bf e} 
& = & \nu_1(q) {\bf r}_{s1} + \nu_2(q) {\bf r}_{s2}, 
\label{r18} \\
\nu_\alpha 
& = & {{\bf v}_\alpha \cdot \hat{\bf n} \over {\bf v}_\alpha \cdot {\bf r}_\alpha},
\label{r98}
\end{eqnarray}
where we have used Eq.~(\ref{r97}) and the fact that 

\begin{equation}
({\bf r}_1 \bbox{\times} {\bf r}_2)\cdot \hat{\bf n} 
= 0
\end{equation}
since $\hat{\bf n}$ is assumed to lie in the plane spanned by ${\bf
r}_1$ and ${\bf r}_2$.  We will now show that Eq.~(\ref{r18}) is valid
not only at noncollinear configurations, but at collinear
configurations as well.  More specifically, we will prove that
$\hat{\bf e}({\bf r}_{s
\alpha})$ is analytic at collinear configurations.

Considering Eq.~(\ref{r18}) in the neighborhood of a collinear
configuration, we first assume that neither ${\bf r}_1$ nor ${\bf
r}_2$ vanishes.  Obviously, to show that $\hat{\bf e} \bbox{(} {\bf r}_{s
\alpha} \bbox{)}$ is analytic, we need only show that $\nu_\alpha({\bf r}_{s
\alpha}) = \nu_\alpha \bbox{(} q({\bf r}_{s \alpha}) \bbox{)}$ is
analytic.  From Lemma~\ref{l1} this amounts to showing that
$\nu_\alpha(q)$ is analytic and even in $q_3$.  That $\nu_\alpha(q)$
is even in $q_3$ is easily proved by inserting Eqs.~(\ref{r99}) and
(\ref{r100}) into Eq.~(\ref{r98}) and using Eqs.~(\ref{r72}) and
(\ref{r73}).  To show that $\nu_\alpha(q)$ is analytic requires the
following observation: The ratio of two functions which are both
analytic at a given point is also analytic at that point provided that
the ratio is not infinite.  (This fact is easily proved by considering
the Taylor series of the two functions.)  As $q_3$ tends toward $0$,
by our previous assumption, ${\bf r}_{s
\alpha}$ does not vanish, nor does  $\hat{\bf n}$.
The quantity ${\bf v}_\alpha$, however, does vanish, but since it
appears to first order in both the numerator and denominator of
Eq.~(\ref{r98}), the ratio is finite at $q_3 = 0$.  Thus,
$\nu_\alpha(q)$ is analytic, and hence $\nu_\alpha({\bf r}_{s \alpha})$
and consequently $\hat{\bf e}({\bf r}_{s \alpha})$ are analytic as well.

If one of the Jacobi vectors, say ${\bf r}_{s1}$, vanishes at the
collinear configuration, the preceding analysis must be modified due
to the appearance of ${\bf r}_1$ in the denominator of Eq.~(\ref{r98}).
In this case, we define a new set of vectors

\begin{eqnarray}
{\bf t}_{s1} & = & {\bf r}_{s1} + {\bf r}_{s2}, \\
{\bf t}_{s2} & = & {\bf r}_{s1} - {\bf r}_{s2},
\end{eqnarray}
and their counterparts ${\bf t}_\alpha$ in the body frame which satisfy

\begin{equation}
{\bf t}_{s \alpha} = {\sf R} {\bf t}_\alpha,
\label{r80}
\end{equation}
analogous to Eq.~(\ref{r20}).  Notice that neither ${\bf t}_1$ nor
${\bf t}_2$ vanishes at the collinear shape since this could only
occur if ${\bf r}_2$ were also to vanish, which only occurs for the
triple collision.  Now, Eqs.~(\ref{r10}) -- (\ref{r100}), (\ref{r18}),
and (\ref{r98}) are all valid if one replaces ${\bf r}_{s
\alpha}$ and ${\bf r}_\alpha$ by ${\bf t}_{s \alpha}$ and ${\bf
t}_\alpha$ respectively.  We may repeat the same line of reasoning as
above to show that the new $\nu_\alpha(q)$, with ${\bf r}_\alpha$
replaced by ${\bf t}_\alpha$, is analytic and even in $q_3$.  Thus
from Lemma~\ref{l1}, the new ${\bf v}_\alpha({\bf r}_{s \alpha})$ and
hence $\hat{\bf e}({\bf r}_{s \alpha})$ are both analytic.  Since
$\hat{\bf n}_1({\bf r}_{s
\alpha})$ and $\hat{\bf n}_2({\bf r}_{s \alpha})$ are analytic, we
finally find that the two mixed coordinates $\hat{e}_i({\bf r}_{s
\alpha}) = \hat{\bf e}({\bf r}_{s \alpha}) \cdot \hat{\bf n}_i({\bf
r}_{s \alpha})$, $i = 1,2$, are analytic functions of the Jacobi
coordinates.

We turn now to the final two mixed coordinates $u_1$ and $u_2$.  They
are defined by Eqs.~(\ref{r23}) and (\ref{r26}), but may also be
expressed as

\begin{eqnarray}
u_1 & = & q_3 \cos \theta 
= \hat{\bf n}_1 \cdot (q_3 {\sf V} {\sf P}_\perp) \hat{\bf n}_1
= \hat{\bf n}_1 \cdot {\sf U}^T (q_3 {\sf R} {\sf P}_\perp ) \hat{\bf n}_1
= \hat{\bf n}_1 \cdot {\sf U}^T {\sf M}  \hat{\bf n}_1, 
\label{r81} \\
u_2 & = & q_3 \sin \theta
= \hat{\bf n}_2 \cdot {\sf U}^T {\sf M}  \hat{\bf n}_1,
\label{r82}
\end{eqnarray}
where 

\begin{equation}
{\sf M} = q_3 {\sf R} {\sf P}_\perp.
\end{equation}
The second equality in Eq.~(\ref{r81}) follows from Eq.~(\ref{r64})
and the orthogonality of $\hat{\bf n}_1$ and $\hat{\bf n}$.  The third
equality follows from Eq.~(\ref{r49}).  

Recall that ${\sf U}({\bf r}_{s \alpha},
\hat{\bf e})$ and  $\hat{\bf e}({\bf r}_{s \alpha})$ are analytic.  Hence, 
${\sf U}( {\bf r}_{s \alpha}) = {\sf U} \bbox{(} {\bf r}_{s \alpha},
\hat{\bf e}({\bf r}_{s \alpha}) \bbox{)}$ is also analytic.  We also
recall that the $\hat{\bf n}_i({\bf r}_{s \alpha})$, $i = 1,2,3,$ are
analytic.  Thus, inspecting Eqs.~(\ref{r81}) and (\ref{r82}), we need
only show that ${\sf M}({\bf r}_{ s\alpha})$ is analytic.  The matrix
${\sf M}$ may be written with the aid of Eq.~(\ref{r10}) as

\begin{eqnarray}
{\sf M}(q, {\bf r}_{s \alpha}) 
& = & {\bf r}_{s1} {\bf p}_1(q)
+ {\bf r}_{s2} {\bf p}_2(q)
+ ({\bf r}_{s1} \bbox{\times} {\bf r}_{s2}) {\bf p}_3^T(q),
\label{r88} \\
{\bf p}_\alpha & = &
{q_3{\sf P}_\perp {\bf v}_\alpha 
\over {\bf v}_\alpha \cdot {\bf r}_\alpha}, \hskip 1cm \alpha = 1,2,
\label{r102} \\
{\bf p}_3 
& = & {q_3{\sf P}_\perp ({\bf r}_{1} \bbox{\times} {\bf r}_{2}) 
\over |{\bf r}_{1} \bbox{\times} {\bf r}_{2}|^2}.
\label{r89}
\end{eqnarray}
We first assume that ${\bf r}_1$ and ${\bf r}_2$ do not vanish at the
collinear shape.  To prove that ${\sf M}({\bf r}_{s \alpha}) = {\sf
M}\bbox{(}q({\bf r}_{s
\alpha}), {\bf r}_{s \alpha}\bbox{)}$ is analytic, we need only show that 
${\bf p}_\alpha({\bf r}_{s \alpha})$, $\alpha = 1,2$, and ${\bf
p}_3({\bf r}_{s \alpha})$ are analytic.  From Lemma~\ref{l1} we must
therefore show that ${\bf p}_\alpha(q)$, $\alpha = 1,2$, and ${\bf
p}_3(q)$ are analytic and even in $q_3$.  Using the definitions
Eqs.~(\ref{r99}), (\ref{r100}), (\ref{r102}), and (\ref{r89}) and the
validity conditions Eqs.~(\ref{r72}) and (\ref{r73}), it is
straightforward to prove that ${\bf p}_\alpha({\bf r}_{s \alpha})$,
$\alpha = 1,2$, and ${\bf p}_3({\bf r}_{s \alpha})$ are even in $q_3$.
As with our proof of the analyticity of $\nu_\alpha(q)$, to prove that
${\bf p}_\alpha(q)$, $\alpha = 1,2$, is analytic all we need to do is
show that the ratio in Eq.~(\ref{r102}) is not infinite when $q_3$
tends toward $0$.  (Both the numerator and denominator are analytic
functions of $q$.)  This fact follows readily from our assumption that
${\bf r}_\alpha$ does not vanish and from the fact that ${\bf
v}_\alpha$, which does vanish at $q_3 = 0$, appears linearly in both
the numerator and denominator.  Proving that ${\bf p}_3(q)$ is
analytic requires a bit more care.  Since both the numerator and the
denominator in Eq.~(\ref{r89}) are analytic functions, we again need
only verify that ${\bf p}_3(q)$ does not blow up at $q_3 = 0$, whence

\begin{equation}
\lim_{q_3 \rightarrow 0} 
{\bf p}_3(q_3)
= \lim_{q_3 \rightarrow 0} 
\left[{2 q_3 \over w_3(q_3)} \right]
\lim_{q_3 \rightarrow 0} 
\left[ {\sf P}_\perp 
\left({{\bf r}_1 \bbox{\times} {\bf r}_2 \over |{\bf r}_1 \bbox{\times} {\bf r}_2 |} \right)
\right] (q_3),
\label{r91}
\end{equation}
where we have used  the definition Eq.~(\ref{r34}) of $w_3$.  We now note

\begin{equation}
\lim_{q_3 \rightarrow 0} {q_3 \over w_3(q_3)}
= {1 \over {\partial w_3 \over \partial q_3} (0)} 
= { \partial q_3 \over \partial w_3}(0),
\label{r90}
\end{equation}
which cannot be infinite since $q_3$ is an analytic function of $w_3$.
(We have used l'Hospital's rule in the second step of
Eq.~(\ref{r90}).)  The second limit in Eq.~(\ref{r91}) cannot be
infinite either, since ${\sf P}_\perp$ is well-defined at $q_3 = 0$
and ${\bf r}_1 \bbox{\times} {\bf r}_2/|{\bf r}_1 \bbox{\times} {\bf
r}_2|$ is a unit vector.  Thus, ${\bf p}_3(q)$ is analytic at $q_3 =
0$.  Hence, we have shown the analyticity of ${\bf p}_{\alpha}({\bf
r}_{s \alpha})$, $\alpha = 1,2$, and ${\bf p}_3({\bf r}_{s \alpha})$
from which follows that analyticity of ${\sf M}({\bf r}_{s \alpha})$
and $u_i({\bf r}_{ s\alpha}), i = 1,2$.  If the assumption that ${\bf
r}_1$ and ${\bf r}_2$ do not vanish at the collinear shape proves to
be false, then we can prove the analyticity of ${\sf M}({\bf r}_{s
\alpha})$ by applying the same trick used earlier of defining the
vectors ${\bf t}_{s
\alpha}$.  We omit the straightforward details of completing this
argument.  This finishes the proof of the compatibility of the mixed
coordinates with the Jacobi coordinates.

We turn now to the eigenspinors $\tau_\mu$ satisfying Eqs.~(\ref{r42})
and (\ref{r78}).  We denote by $(L_{si})_{m m'}$, $i = x,y,z$, $m,m' =
-\ell, ...,
\ell$, the components of the space-referred angular momentum operator
${L}_{si}$ with respect to the $L_{sz}$ eigenbasis.  The matrices
$L_{si}$ transform under a rotation ${\sf Q}\in SO(3)$ as a vector
operator,

\begin{equation}
\sum_j {\sf Q}_{ji} L_{sj} 
= {\cal D}^\ell({\sf Q}) L_{si}  {\cal D}^{\ell \dagger}({\sf Q}).
\end{equation}
We denote by $(L_{i})_{k k'}$, $i=x,y,z$, $k,k' = -\ell, ..., \ell$,
the components of the body-referred angular momentum operator $L_i$
with respect to the $L_z$ eigenbasis.  The components of the two
operators $L_{si}$ and $L_i$, with respect to their respective bases,
are related by $(L_{i})_{k k'} = (L_{si})_{k' k} = (L_{si})_{k k'}^*$.
(See Ref.~\cite{Littlejohn97}, Sect. 4.H for a derivation.)  Thus, the
matrices ${L}_{i}$ satisfy

\begin{equation}
\sum_j {\sf Q}_{ji} {L}_j 
= {\cal D}^{\ell *}({\sf Q}) 
{L}_i  {\cal D}^{\ell T}({\sf Q}).
\label{r45}
\end{equation}
In view of Eqs.~(\ref{r43}) and (\ref{r45}), we have 

\begin{equation}
\hat{\bf n} \cdot {\bf L}
= \hat{\bf z} \cdot {\sf W}^T {\bf L}
= {\cal D}^{\ell *}({\sf W}) L_z {\cal D}^{\ell T}({\sf W}).
\label{r44}
\end{equation}
From Eq.~(\ref{r44}) we see that $\tau_\mu$ can be expressed as 

\begin{equation}
(\tau_\mu)_k 
= e^{i \sigma}{\cal D}^\ell_{\mu k}  \left({\sf W}^T\right),
\end{equation}
where $\sigma(q)$ is a phase factor which must be analytic and even in
$q_3$ in order to guarantee the same properties for $(\tau_\mu)_k(q)$.
This result, together with Eqs.~(\ref{r46}) and (\ref{r79}), yields

\begin{eqnarray}
\chi_\mu^\ell 
& = & \sum_k e^{-i\sigma}{\cal D}^{\ell *}_{\mu k}({\sf W}^T)\chi^\ell_k, 
\label{r48} \\
\chi_k^\ell 
& = & \sum_\mu e^{i \sigma} {\cal D}^{\ell *}_{k \mu}({\sf W})\chi^\ell_\mu.
\label{r68}
\end{eqnarray}
We may use Eq.~(\ref{r68}) to rewrite Eq.~(\ref{r7}) as

\begin{equation}
\Psi^\ell_m 
= e^{i \sigma} \sum_\mu {\cal D}^{\ell *}_{m \mu}({\sf R} {\sf W}) \chi^\ell_\mu
= e^{i \sigma} \sum_\mu {\cal D}^{\ell *}_{m \mu}({\sf U} {\sf V} {\sf W}) \chi^\ell_\mu,
\label{r69}
\end{equation}
where we have used Eq.~(\ref{r49}).

We now assume that $\Psi^\ell_m({\bf r}_{s \alpha})$ is analytic,
which implies, as in part (i), that both $\chi^\ell_k(q)$ and
$\chi^\ell_\mu(q)$ are analytic, and hence $\chi^\ell_\mu(q)$ has a
Taylor series

\begin{equation}
\chi^\ell_\mu(q_1, q_2, q_3) 
= \sum_{n = 0}^\infty b_{n \mu}(q_1,q_2) q_3^n,
\label{r47}
\end{equation}
where $b_{n \mu}(q_1, q_2)$ is analytic.  We must now only show that
the appropriate coefficients vanish in the Taylor series
Eq.~(\ref{r47}) to produce the Taylor series Eq.~(\ref{r12}).  Since
$\sigma(q)$ is analytic and even in $q_3$, Lemma~\ref{l1} shows that
$\sigma({\bf r}_{s \alpha}) = \sigma \bbox{(} q({\bf r}_{s \alpha})
\bbox{)}$ is analytic.  Therefore the function $\tilde{\Psi}^\ell_m$
defined by

\begin{equation}
\tilde{\Psi}^\ell_m(q_1, q_2, u_1, u_2) 
= e^{-i\sigma(q_1, q_2, u_1, u_2)}
\Psi^\ell_m(q_1, q_2, u_1, u_2, \hat{\bf e} = \hat{\bf z}),
\label{r95}
\end{equation}
is analytic.  Since $\hat{\bf e} = \hat{\bf z}$, we see from
Eq.~(\ref{r84}) that the matrix product ${\sf U} {\sf V} {\sf W}$
appearing in Eq.~(\ref{r69}) is equal to ${\sf W}^T{\sf V} {\sf W}$.
From Eqs.~(\ref{r64}) and (\ref{r43}) it is evident that ${\sf
W}^T{\sf V} {\sf W} \hat{\bf z} =
\hat{\bf z}$, and hence ${\sf W}^T{\sf V} {\sf W}$ is a rotation by
$\theta$ about the $z$-axis, where $\theta$ is the rotation angle of
${\sf V}$.  Therefore,

\begin{equation}
{\cal D}^{\ell *}_{m
\mu}({\sf W}^T {\sf V} {\sf W}) 
= \delta_{m \mu} \exp(i m \theta).  
\label{r96}
\end{equation}
Thus, Eqs.~(\ref{r69}) and (\ref{r95}) combine to yield

\begin{equation}
\tilde{\Psi}^\ell_m( u_1, u_2) 
= e^{im\theta} \chi^\ell_m(q_3),
\label{r94}
\end{equation}
where we have dropped the explicit dependence on $q_1$ and $q_2$.
From this equation and the fact that $(\theta, q_3)$ are the usual
polar coordinates with respect to the Cartesian coordinates $(u_1,
u_2)$, we may apply Theorem \ref{t2}.  In particular, since
$\tilde{\Psi}^\ell_m(u_1, u_2)$ is analytic, $\chi^\ell_m(q_3)$ has
the Taylor series given in Eq.~(\ref{r15}).  This implies that the
appropriate coefficients in Eq.~(\ref{r47}) vanish.

Next assume that $\chi^\ell_\mu(q)$, and hence $\chi^\ell_k(q)$, is
analytic and that $\chi^\ell_\mu(q)$ has the Taylor series given by
Eq.~(\ref{r12}).  Then from Theorem \ref{t2}, the function
$\tilde{\Psi}^\ell_m(u_1, u_2)$ given in Eq.~(\ref{r94}) is analytic
in $u_1$ and $u_2$ and based on the analyticity of the coefficients
$a_{\mu n}(q_1, q_2)$, $\tilde{\Psi}^\ell_m(u_1, u_2, q_1, q_2)$ is
analytic in $q_1$ and $q_2$ as well.  We now rewrite Eq.~(\ref{r69})
as

\begin{equation}
\Psi^\ell_m 
= e^{i \sigma} 
\sum_\mu {\cal D}^{\ell *}_{m \mu}({\sf U} {\sf W}) e^{i \mu \theta}\chi^\ell_\mu
= e^{i \sigma} 
\sum_\mu {\cal D}^{\ell *}_{m \mu}({\sf U} {\sf W}) \tilde{\Psi}^\ell_\mu,
\end{equation}
where we have used Eq.~(\ref{r96}) in the first equality and
Eq.~(\ref{r94}) in the second.  We have already shown that $\sigma$,
${\sf U}$, ${\sf W}$, and $\tilde{\Psi}^\ell_\mu$ are analytic
functions of the mixed coordinates.  Thus, $\Psi^\ell_m$ is an
analytic function of the mixed coordinates and hence of the Jacobi
coordinates as well. ${\cal QED}$.

\section{Conclusions}
\label{s5}

Assuming valid conventions, Theorem~\ref{t1} solves the problem of
determining whether an internal wave function for the three-body
problem is associated with an analytic external wave function.  The
criteria presented are both necessary and sufficient.  We now propose
several useful extensions of the present work which we plan to
consider in future publications.

First, we observe that Theorem \ref{t1} says nothing about the
properties of the external wave function in the neighborhood of body
frame singularities.  Our justification for ignoring these
singularities at present is that they may always be moved away from
the region of interest by a change of body frame; they may even be
moved into the unphysical region of shape space.  However, in practice
it is not always convenient to change body frames or to use a body
frame which places the singularities in the unphysical region.
Therefore, one should like to have a set of criteria on the internal
wave function which guarantees the analyticity of the external wave
function in the neighborhood of a body frame singularity.

A second desirable extension of the present work would be to develop a
set of criteria applicable at the three-body collision.  Such an
analysis would almost certainly involve the analysis of body frame
singularities mentioned above.  This is because the string (or
strings) of body frame singularities emanates from the three-body
collision and no body frame avoids singularities at the three-body
collision.

As mentioned in the introduction, the external wave function for
three-body problems is not necessarily analytic at all points, the
collisional configurations of Coulomb problems being an obvious
example.  Yet another useful extension of the present work would be to
develop criteria on the internal wave function which capture such
singular behavior of the external wave function.

Finally, it is natural to try and extend our analysis to four or more
bodies.  The case of four bodies is certainly more challenging than
that of three bodies.  For the three-body problem, we can always pick
valid reference orientations ${\bf r}_\alpha(q)$ which are well
defined analytic functions in the neighborhood of an arbitrary
collinear shape (except the three-body collision).  For the four-body
problem, however, any reference orientation ${\bf r}_\alpha(q)$,
$\alpha = 1,2,3$, we choose will not be analytic at any collinear
shape.  This is because all body frames for the four body problem have
singular surfaces emanating from all collinear shapes \cite{Littlejohn98b}.
Extending the analysis of this paper to the four body problem
will necessarily require a deeper consideration of frame singularities.

\section{Acknowledgments}
The authors greatly appreciate the thorough reading and critique given the
manuscript by Dr. M. M\"uller.  The research in this paper was
supported by the U. S. Department of Energy under Contract
No. DE-AC03-76SF00098.

\appendix
\section{Facts Concerning $SO(2)$}

We collect some basic facts concerning $SO(2)$ which are necessary for
the proof of Theorem \ref{t2}.  All irreducible representations
(irreps) of $SO(2)$ are one-dimensional and may be labeled by an
integer $m$, with $-\infty < m < \infty$.  The components of a vector
in an invariant one-dimensional carrier space labeled by $m$ are
multiplied by $\exp( - i m
\theta)$ when rotated by $\theta$.

The fundamental representation of $SO(2)$, that is the representation
by $2 \times 2$ real orthogonal matrices, decomposes into the direct
sum of the $m = \pm 1$ irreps, which we denote
$+1 \oplus -1$. The two irreducible carrier spaces are spanned by the
vectors ${\bf e}_\pm = (1, \pm i)^T$.  The $d$-fold tensor product of
the fundamental representation contains (with various multiplicities)
the irreps $d$, $d-2$, $d - 4$, ..., $-(d-4)$,
$-(d-2)$, $-d$.  However, exactly one irrep for
each allowed value of $m$ is totally symmetric.  That is, the fully
symmetrized tensor product of the fundamental representation of
$SO(2)$ decomposes as

\begin{equation}
{\cal S} \bigotimes_d ( + 1 \oplus -1 )
= d \oplus (d-2) \oplus (d-4) \oplus \cdots \oplus -(d-4) \oplus -(d-2) \oplus -d,
\label{r4}
\end{equation}
where $\otimes$ denotes the tensor product and ${\cal S}$ denotes the
total symmetrization operator.  The irrep labeled by $m$ is spanned by
a totally symmetric rank $d$ tensor $t^d_m$ given by

\begin{equation}
t^d_m 
= {\cal S} ( \underbrace{{\bf e}_+ \otimes \cdots \otimes {\bf e}_+}_{ u \mbox{ times}} 
\otimes \underbrace{{\bf e}_- \otimes \cdots \otimes {\bf e}_-}_{v \mbox{ times}}),
\label{r17}
\end{equation}
where $m = u-v$ and $d = u + v$.  Explicitly, $t^d_m$ transforms as

\begin{equation}
\sum_{k_1...k_d} Q_{j_1 k_1} ... Q_{j_d k_d} (t^d_m)_{k_1 ... k_d} 
= e^{ - i m \alpha} (t^d_m)_{j_1 ... j_d}, 
\label{r16}
\end{equation}
where $(t^d_m)_{j_1 ... j_d}$ are the components of $t^d_m$.  A final
fact we need, which follows readily from Eq.~(\ref{r17}), is that

\begin{equation}
(t^d_m)_{x ... x} = 1.
\label{r38}
\end{equation}

\end{document}